\documentclass[12pt]{article}

\usepackage{amssymb,graphicx,amsmath}

\usepackage{epsf}
\usepackage{graphicx,epsfig}
\usepackage{amsfonts}
\usepackage{amssymb}

\def\bk{{\bf k}}

\def\bx{{\bf x}}

\def\CO{{\cal O}}

\def\CR{{\cal R}}
\def\CG{{\cal G}}

\def\mpl{M_p}
\def\half{\frac{1}{2}}

\makeatletter
\renewcommand\section{\@startsection {section}{1}{\z@}%
                                 {-3.5ex \@plus -1ex \@minus -.2ex}
                                   {2.3ex \@plus.2ex}%
                                   {\normalfont\large\bfseries}}
\renewcommand\subsection{\@startsection{subsection}{2}{\z@}%
                                   {-3.25ex\@plus -1ex \@minus -.2ex}%
                                     {1.5ex \@plus .2ex}%
                                     {\normalfont\bfseries}}
\renewcommand\subsubsection{\@startsection{subsubsection}{3}{\z@}%
                                   {-3.25ex\@plus -1ex \@minus -.2ex}%
                                     {1.5ex \@plus .2ex}%
                                     {\normalfont\itshape}}
\makeatother

\newcommand{\Letter}{
\setlength{\textwidth}{16.5cm}
   \setlength{\textheight}{22.6cm}
    \hoffset=-0.5in
\voffset=-2.1cm }

\Letter

\begin{document}
\newcommand{\be}{\begin{equation}}
\newcommand{\ee}{\end{equation}}
\newcommand{\bea}{\begin{eqnarray}}
\newcommand{\eea}{\end{eqnarray}}
\newcommand{\barr}{\begin{array}}
\newcommand{\earr}{\end{array}}
\newcommand{\myfigure}[2]{\resizebox{#1}{!}{\includegraphics{#2}}}

\thispagestyle{empty}
\begin{flushright}
\parbox[t]{1.8in}{
MIT-TP-3923\\
CAS-KITPC/ITP-048}
\end{flushright}

\vspace*{0.3in}

\begin{center}
{\large \bf Generation and Characterization of Large Non-Gaussianities  \\
\vspace{0.3cm}
in Single Field Inflation 
}

\vspace*{0.5in} {Xingang Chen${}^1$, Richard Easther${}^2$ and Eugene A. Lim${}^{3}$ } 
\\[.3in]
${}^1$ Center for Theoretical Physics, \\
Massachusetts Institute of
Technology, Cambridge, MA 02139 \\
${}^2$   Department of Physics, Yale University, New Haven, CT 06511\\
${}^3$ ISCAP and Physics Department, Columbia University, New York, NY 10027
 \\[0.3in]
\end{center}

\begin{center}
{\bf
Abstract
}
\end{center}
\noindent

Inflation driven by a single, minimally coupled, slowly rolling field
 generically yields a negligible primordial
non-Gaussianity.  We discuss  two distinct
mechanisms by which a non-trivial potential can generate large
non-Gaussianities.  Firstly, if the inflaton traverses a  feature in the potential, or if the inflationary phase is short enough so that initial transient contributions to the background dynamics have not been erased,  modes near horizon-crossing can acquire significant non-Gaussianities. Secondly, potentials with small-scale structure may induce significant non-Gaussianities while the relevant
modes are deep inside the horizon.    The first case includes the ``step'' potential we previously analyzed while the second ``resonance'' case is novel.   We
derive analytic approximations for the 3-point terms
generated by both mechanisms written as 
products of
functions of the three individual momenta,  permitting  the use of efficient analysis algorithms.     
Finally,  we present a significantly improved approach to regularizing and numerically evaluating the integrals that contribute to the  3-point function.
\vfill 

\newpage
\setcounter{page}{1}

\tableofcontents


\section{Introduction}

The primordial power spectrum or 2-point correlation function  of the temperature anisot\-ropies in the Cosmic Microwave Background (CMB) is well-measured out to large multipoles. The higher
moments of a distribution are, in general, independent of the 2-point function, but the CMB anisotropies are at least approximately Gaussian.   Theoretically, we know that non-Gaussian component to the CMB will always be induced by  non-linear gravitational couplings between modes after they reenter the horizon
\cite{Pyne:1995bs}  while  single
field slow-roll inflation yields a primordial  non-Gaussian
component roughly 100 times smaller than that induced by gravitational
mode-couplings  \cite{Acquaviva:2002ud,Maldacena:2002vr}.   
The two terms are additive, so recovering this latter 
primordial signal is next to impossible, even before contending with
cosmic variance.  Multi-field models may generate larger
non-Gaussianities
\cite{Rigopoulos:2005us,Seery:2005gb,Vernizzi:2006ve,Battefeld:2006sz,Yokoyama:2007uu},
but this is by no means a generic property of these scenarios.    
Consequently, the detection of a primordial 3-point function would immediately
falsify a very large class of inflationary models. 
Conversely, non-slow-roll models with 
higher order derivative terms, such as DBI inflation
\cite{Silverstein:2003hf,Alishahiha:2004eh,Chen:2004gc,Chen:2005ad}
and k-inflation \cite{ArmendarizPicon:1999rj,Garriga:1999vw}, do
typically generate large non-Gaussianities  \cite{Chen:2006nt}. Further
references, include more complicated multi-field, non-local or ghost
theories, can be found in
Refs.~\cite{Li:2008qc,Bean:2008ga,Lyth:2002my,Chen:2007gd,Gupta:2002kn,Moss:2007cv,Barnaby:2007yb,ArkaniHamed:2003uz,Cheung:2007st}.
In this paper we investigate simple models -- in the sense that they
are driven by a single, minimally coupled scalar field with a
canonical kinetic term -- which generate substantial
non-Gaussianities.

Constraining the non-Gaussian signal in a CMB dataset is 
a highly non-trivial problem.  Firstly, it depends on the choice of estimators. At the
moment, only two concrete estimators have been constructed: the
$f_{NL}^{local}$ and $f_{NL}^{equil}$ forms 
\cite{Komatsu:2001rj,Komatsu:2003iq,Babich:2004gb,Spergel:2006hy,Creminelli:2006gc,Creminelli:2006rz,Yadav:2007ny}, and both are
scale-invariant. 
The recent claim of a detection of a non-zero 3-point function \cite{Yadav:2007yy} in the WMAP 3-year data \cite{Spergel:2006hy} relies on the estimator developed in Ref.~\cite{Yadav:2007ny,Creminelli:2006gc,Komatsu:2003iq} which is
of the ``local'' form  \cite{Salopek:1990jq,Komatsu:2001rj}.
However non-Gaussianities that have strong scale dependence 
are not well-described by $f_{NL}^{local}$ and $f_{NL}^{equil}$, and will require  scale-dependent  estimators, based
on theoretically motivated 
ansatzen 
for the primordial
non-Gaussianities.  In addition to computing the primordial 3-point term, comparing the predictions of a specific inflationary model to the CMB requires us to evolve this signal through to recombination, and to project it onto the sphere on the sky, which is a convolution of the
primordial 3-point term with 3 $l$-valued spherical Bessel functions. In general, this process is computationally expensive but  simplifies dramatically if the 3-point
function has special
algebraic properties \cite{Smith:2006ud,Fergusson:2006pr}.  

Using Maldacena's elegant formalism \cite{Maldacena:2002vr}, the primordial
3-point curvature correlation function is 
computed via a set of integrals (over time) of
products of three mode functions (or their derivatives) and slow
roll parameters.  Looking at
these integrals, we identify two mechanisms which can create
substantial non-Gaussianities.  The first class consists of potentials
with a localized violation of slow roll. This can take the form of  a small localized feature, including the step models
\cite{Adams:2001vc,Komatsu:2003fd} 
whose 3-point term was first accurately computed
by the present authors in \cite{Chen:2006xj}. In addition models with a short inflationary phase can have initial transients in their dynamics, which will not be fully erased before the longest modes leave the horizon.  In these cases, the 3-point term for modes which are leaving the horizon during the violation of slow roll can be magnified by three orders of magnitude, without ruining the fit to the 2-point function.  The
second class arises when  a small ripple is superimposed on
top of an otherwise smooth potential.  This induces a ``resonance''
inside one of these integrals, giving the 3-point function a
substantial amplitude {\em before} the modes cross the
horizon. In the former case the 2-point function may be
substantially
modified; in the second case the modification of the 
2-point function is very small, even though the 3-point function is
large. The latter mechanism has not been described previously and 
the seemingly contrived field theoretic potential may in fact arise
naturally in brane inflation models \cite{Bean:2008na}.
 
With this information in hand, we construct rough analytic approximations to
the corresponding 3-point terms.  These expressions
have the factorizable form required by  \cite{Smith:2006ud}, which
means that we can efficiently compute constraints on step potentials
or similar models from CMB data, although at this point we are only interested in a qualitative match to the numerically computed 3-point term.    
  We check our semi-analytic estimates for the 3-point function using an improved version of the code described in
\cite{Chen:2006xj}, which has a  much cleaner scheme for
removing numerical divergences in the 3-point integrals. The
integrands are products of large, rapidly oscillating
terms. Analytically, these are finite, but their {\em numerical\/}
evaluation is non-trivial, and we show how 
to transform them into an explicitly finite form before the numerical evaluation is carried out.   From a  practical perspective, this means we can compute the 3-point function for ``triangles'' which contain two very different scales.  

The paper is organized as follows. In Section \ref{sect:MW} we review
the computation of  the 3-point function, and in Section  \ref{sect:NumInt} we describe the numerical algorithm.  
Sections \ref{sect:horizon} and \ref{sect:subhorizon}  discuss the two
distinct  
\emph{mechanisms} for generating large primordial non-Gaussianities
within minimally coupled  single scalar field inflation,  and  we derive
the approximate analytic ansatzen for the 3-point function. We then
use our numerical methods to compute the exact 3-point
function for these models, and show that these approximations yield   fair
representations of the underlying non-Gaussianities. In Section
\ref{sect:Conclusions} we summarize and discuss our results and plans 
for future work.   We follow the 
notational conventions of \cite{Chen:2006xj} and  set the reduced
Planck mass $M_p$ to unity except when presenting final results or defining parameter values.
 
\section{Single scalar field inflation models and Maldacena's
formalism} 
\label{sect:MW}
 Consider scalar field
inflation with an arbitrary potential 
\begin{equation}
S = \int dx^4 \sqrt{g}\left[ \frac{R}{2} - \frac{1}{2}(\partial
\phi)^2-V(\phi)\right] \label{eqn:scalarfieldaction}. 
\end{equation}
During inflation, spacetime is described by the
Friedman-Robertson-Walker metric
\begin{equation}
ds^2 = - dt^2 + a^2(t) (dr^2 + r^2 d\Omega) = a^2 (-d\tau^2 + dr^2 +
r^2 d\Omega) ~,
\end{equation}
and the conformal time $\tau$ runs from $-\infty$ to $0$. Dots denote
derivatives by cosmic time $t$ while primes denote
derivatives with respect to $\tau$. 

The evolution of the single scalar field is described by  
\begin{equation}
\phi''  +2 \frac{a'}{a}\phi' + \frac{1}{a^2}\frac{dV}{d\phi}=0 ~,
\label{eqn:scalareom}
\end{equation}
while the scale factor obeys the Friedmann equation
\begin{equation}
H = \frac{\dot{a}}{a}=  \frac{1}{3}\left[\frac{1}{2}(\partial \phi)^2
+ V(\phi)\right] ~.
\label{eqn:hubble}
\end{equation}
The solution of the scalar field to these coupled set of 
equations is the trajectory of the scalar field in phase space
$(\phi(t),\dot{\phi}(t))$. We can also describe this trajectory with
the  slow-roll parameters \cite{Salopek:1990re}
\begin{eqnarray}
\epsilon &=& -\frac{1}{a}\frac{H'}{H^2} ~,
\label{eqn:epsilon}  \\
\eta &=& \frac{\dot \epsilon}{\epsilon H} =
\frac{1}{a}\left[\frac{H''}{H H'}-a-2\frac{H'}{H^2}\right] ~.
\label{eqn:eta}
\end{eqnarray}
Slow-roll inflaton
occurs  when $|\epsilon|\ll 1$ and $|\eta|\ll 1$. 
 
Since we are interested in models with complicated potentials, we will
need to solve their perturbation equations numerically, before deriving analytical approximations to the 3-point functions.
Moreover,  we will need to track the perturbations while they are inside  the horizon, and thus cannot resort to the   large wavelength approximation
\cite{Yokoyama:2007uu,Sasaki:1995aw,Byrnes:2007tm,Lyth:2005fi}.  Instead we use the approach introduced by
Maldacena \cite{Maldacena:2002vr}, in what follows we use the notation of both this paper and Ref. \cite{Weinberg:2005vy}. 

We first  split the Hamiltonian into its
quadratic component $H_0$ and its higher order interaction component
$H_I$,
\begin{equation}
H[\delta \phi, \delta g_{\mu\nu}] 
= H_0[\delta \phi^I, \delta g_{\mu\nu}^I] + 
H_I[\delta \phi^I, \delta g_{\mu\nu}^I] ~.
\end{equation}
The superscript $I$ signifies that these modes are evolved using the
linear (i.e.~free field) equations of motion. Since we are interested
only in the 3-point correlation functions, $H_I$ contains terms up to
third order in linear variables. 

One can then show (see the Appendix of \cite{Weinberg:2005vy}) that any
 given combination of product of fields  $Q$, can be evolved
 by a simple unitary transform 
\begin{equation}
Q(t) = \left[\bar{T}\exp\left(i\int^t_{t_0}H_I(t) dt\right)\right]
Q^I(t)\left[T\exp\left(-i\int^t_{t_0}H_I(t)dt\right)\right] \,
. \label{eqn:generalQ}
\end{equation}
Here $\bar{T}$ and $T$ refer to anti-time-ordering and time-ordering,
but since we  have only third order terms, the time ordering is not a
factor in the equation. Using Eq.~(\ref{eqn:generalQ})
\cite{Maldacena:2002vr}, the 3-point correlation
function for the Bardeen curvature $\zeta$ \cite{Bardeen:1980kt}
is
\begin{equation} 
\langle\zeta(\tau,\textbf{k}_1)\zeta(\tau,\textbf{k}_2)\zeta(\tau,\textbf{k}_3)\rangle=
-i\int_{\tau_0}^{\tau} d\tau' ~ a ~ \langle
[
\zeta(\tau,\textbf{k}_1)\zeta(\tau,\textbf{k}_2)\zeta(\tau,\textbf{k}_3),{H}_{I}(\tau')]
\rangle ~. \label{eqn:ehrenfest}
\end{equation}
This is well-suited to numerical calculations, as we now simply need to solve
for the Fourier mode of the linear order perturbation in order to
compute the relevant integrals. 

The main advantage of Maldacena's approach is the usage of the ADM
formalism, where the constraint equations can be conveniently solved and
interaction terms expanded.
The computation of $H_I$ for a minimally coupled 
single field inflationary model
was performed in detail in Ref.~\cite{Maldacena:2002vr}.
It has the general form
\begin{equation}
H_I = \int dx^3 \sum_i a^2 g_i(\epsilon,\eta,\eta') \xi_1 \xi_2 \xi_3 ~,
\label{eqn:HIexpansion}
\end{equation}
where $\xi$ denotes either $\zeta$, $\zeta'$ or $\partial \zeta$, and
$g_i$ are functions of the slow-roll parameters $\epsilon(t)$ and
$\eta(t)$. We see that the coupling strength of the interaction
Hamiltonian depends on the dynamics of the background encoded by the
slow-roll parameters. We emphasize that this equation is \emph{exact} to
all orders in slow-roll parameters, and \emph{does not depend on the
slow-roll conditions}. In other words, even if we violate the slow
roll conditions these equations still hold. 
In order to
compute the 3-point function, we must evaluate the set of integrals
listed in Appendix \ref{section:3pts}.  It is now clear why standard single field slow
roll inflation does not generate large non-Gaussianities: the
interaction couplings are functions of the slow-roll parameters, and
hence are small by construction. However, large non-Gaussianities are
possible if these couplings behave in a non-trivial way while keeping
the viability of the power spectrum.

\section{Numerical integration} \label{sect:NumInt}

Our goal is to integrate Eqs.~(\ref{eqn:ehrenfest}) and
(\ref{eqn:HIexpansion}) numerically. We first numerically solve
the background and linear order equations of motion for the fields
using the free field Hamiltonian -- to do this we solve the relevant
equations of motion using a standard Runge-Kutta 6th order integrator.
We then plug the solution into Eq.~(\ref{eqn:ehrenfest}) and integrate
them from $-\infty < \tau <0$ to obtain the 3-point function. 

The integrals (see Appendix \ref{section:3pts} for the explicit formulas) possess the generic form
\begin{equation}
I\propto \Re \left[\prod_i
u_i(\tau_{end})\int_{-\infty}^{\tau_{end}}d\tau a^2 g(\epsilon,\eta,\eta')
\xi_1(\tau)\xi_2(\tau)\xi_3(\tau)  \right] \label{eqn:Itype1}
\end{equation}
where $\xi$ can denote $\zeta(\vec{k})$ or
$d\zeta(\vec{k})/d\tau$, and $g$ is some function of the slow roll parameters which differs from term to term. These integrals are formally convergent in
the limit  $\tau\rightarrow-\infty$, but cutting them off at a finite
value of $\tau$ exposes the oscillatory nature of the integrand whose
amplitude blows up rapidly as $\tau$ grows large and negative,
introducing a spurious  contribution of
${\cal{O}}(1)$. Physically, when the modes are well within the
horizon, they  oscillate rapidly compared to the rate of change of the
interaction terms, so the contribution to the integrals almost
cancels.  We can see this by rotating the integrals $\tau\rightarrow
\tau(1+i\epsilon)$ into the imaginary plane, giving the oscillatory
part of the integral a damping term at large negative $\tau$
\cite{Maldacena:2002vr,Seery:2005wm,Chen:2006nt}.

In \cite{Chen:2006xj}, we added in a damping factor $e^{-\beta \tau}$ 
by hand  to   regulate the integrals, but this tends to systematically underestimate the resulting integrals, and $\beta$ needs to be chosen carefully to balance accuracy and computational efficiency.   However, we can  regulate these integrals analytically, so that the numerical evaluation involves an explicitly finite integrand. We start by splitting the integral in Eq. (\ref{eqn:Itype1})  into two parts 
\begin{equation}
I = \int_{-\infty}^{\tau_{end}} = \int_{-\infty}^{\tau_0} + \int_{\tau_0}^{\tau_{end}} = I_1 + I_2 \,.
\end{equation}
Here $\tau_0$  is an arbitrary time when all three modes are well
inside the horizon, and $\tau_{end}$ corresponds to a moment long
after horizon exit.   It is straightforward to numerically evaluate
$I_2$, but $I_1$ suffers from the cut-off dependence we mentioned
earlier.  

When all three modes are well inside the horizon, their phase and
amplitude is  well-described by the WKB approximation \cite{Bender}\footnote{The 2-point function is computed with WKB methods in
\cite{Martin:2002vn}.  Here we only need the mode evolution inside
the horizon, while a full treatment  requires matching across the
moment of horizon crossing, using standard turning point
techniques.}.
\begin{equation}
v_k \approx \frac{1}{\sqrt{2\alpha(k)}}\exp\left[i\int
  \sqrt{\alpha(k)} d\tau\right] +\mathrm{c.c} ~,
\end{equation}
where $\alpha(k) = k^2+z''/z$. Deep within the horizon, $k^2\gg z''/z$, the modes would not see the curvature term and hence it will propagate as a plane wave $v_k \propto \exp[i k\tau]/\sqrt{k}$.
 Specializing to the case $\xi = \zeta$ for now
\begin{equation}
I_1 = \int_{-\infty}^{\tau_0} d \tau \theta(\tau) \frac{1}{\sqrt{8k_1 k_2 k_3}} e^{i K(\tau-\tau_0)} \label{eqn:I2}
\end{equation}
where $K\equiv k_1+k_2+k_3$ and we set
\begin{equation}
v(\tau_0) = \frac{1}{\sqrt{2k}} ~, \qquad 
v'(\tau_0) = \frac{i k}{\sqrt{2k}},
\end{equation}
and $\theta$ is some function of $\tau$ given by
\begin{equation}
\theta(\tau) = \frac{a^2}{z^3}g(\epsilon,\eta,\eta').
\end{equation}
In general, $\theta$ diverges as  $\tau\rightarrow -\infty$, but there is
always some finite value of $p$ such that $\theta(\tau) \tau^p \rightarrow
0$ as $\tau\rightarrow -\infty$.

\begin{figure}[t]
\myfigure{5in}{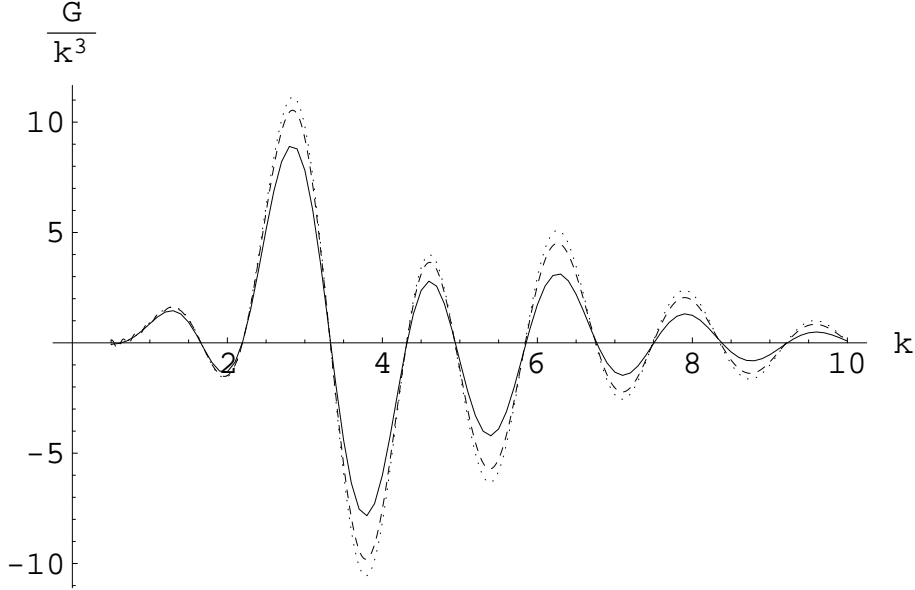}
\caption{This plot compares the 3-point correlation function, computed using the 
$\beta$ described in \cite{Chen:2006xj}  and the boundary regulator described in this section, in the case of the step model
[Sec.~\ref{subsect:sharpfeatures} with model parameters $(c,d,\phi_s)
= (0.0018,0.022M_p,14.84M_p)$].  We plot the dimensionless variable
$G/k^3$ defined in Eq.~(\ref{eqn:g}) for the equilateral case. The
solid and the dashed lines are results obtained  $\beta=0.01$ and $\beta=0.005$ respectively, while the
dotted line is the result obtained from using the boundary
regulator. The value of $\beta$ is chosen such that it gives optimal
results at around $k=1$; if $\beta$ is too small the
early time oscillation will not be suppressed while if $\beta$ is too large it will over-suppress the high $k$ values. The boundary
regulator does not suffer from this arbitrariness, and matches the limit found when $\beta\rightarrow0$. } 
\label{fig:comparisonplot}
\end{figure}

Now integrate Eq. (\ref{eqn:I2}) by parts 
\begin{equation}
I_1 = \frac{1}{\sqrt{8k_1 k_2 k_3}}   \frac{-i}{K}\left. \theta(\tau)
e^{i K(\tau-\tau_0)}\right|_{-\infty}^{\tau_0} -
\int_{-\infty}^{\tau_0} d\tau \frac{d\theta}{d\tau}\frac{1}{\sqrt{8k_1 k_2
k_3}}\frac{-i}{K}e^{i K(\tau-\tau_0)}. 
\end{equation}
The boundary term at $-\infty$ is apparently divergent, but if we use
the same $\epsilon$ rotated contour as before, so $-\tau \rightarrow
-\infty(1+i\epsilon)$ the term vanishes for any finite value of
$\epsilon$.  The remaining integral  is now \emph{more} convergent at
large negative $\tau$ since the integrand has picked up a factor $\propto
\tau^{-1}$.  Integrating by parts  a second time we find 
\begin{equation}
I_1 = \frac{1}{\sqrt{8k_1 k_2
k_3}}\left[\frac{-i}{K}\theta(\tau_0)-\left(\frac{-i}{K}\right)^2\frac{d\theta}{d\tau}(\tau_0)\right]+\int_{-\infty}^{\tau_0}
d\tau \frac{d^2 \theta}{d\tau^2}\frac{1}{\sqrt{8k_1 k_2
k_3}}\left(\frac{-i}{K}\right)^2e^{i K(\tau-\tau_0)}. 
\end{equation}
The resulting integral is now convergent for any $F$ we encounter in this work, and
can be evaluated efficiently. We could repeat this
process to further speed the convergence of the remaining integral,
but in practice this is not necessary.    We illustrate the
effectiveness of this ``boundary regulator'' in Figure
\ref{fig:comparisonplot}.    
 
\section{Horizon scale generation of non-Gaussianities} 
\label{sect:horizon}

In this section, we explore models of inflation where the
non-Gaussianities are generated when the modes cross the
horizon. Typically these models require a violation of slow-roll at
some fixed physical scale. All modes experience a temporary boost
in their coupling strengths courtesy of this violation, but only modes
exiting the horizon as the violation
occurs receive a boost in their non-Gaussian signatures. Modes deep within the horizon are still rapidly
oscillating, and the net contribution to their non-Gaussianities
cancel.   The violation of slow roll can have two origins. The first is a potential with a localized  feature, as discussed in   \cite{Chen:2006xj},   and we now generalize this  analysis.  Secondly,  if the duration of inflation is such that initial transients in the dynamics have not been erased before observable modes leave the horizon   \cite{Contaldi:2003zv} we again find a significantly boosted 3-point function at these scales, even though the {\em potential\/} is smooth.

\subsection{Features in the inflationary potential} \label{subsect:sharpfeatures}

Consider a small step in the  inflaton potential  \cite{Adams:2001vc}. 
In the limit that the step is a genuine discontinuity, the change in the potential energy across the step would be entirely converted into the inflaton's kinetic energy, which is then damped away. For realistic models  $\Delta V /V < 0.01$ so $\dot{\phi}^2/2 \ll V$ across the step, and  $\epsilon \ll 1$. Recall that we are working in the Hubble slow roll formalism -- if we had defined $\epsilon \propto (V_{,\phi}/V)^2$ this quantity can become large across the step.  Further,  $\eta$ is the rate of change of $\epsilon$, so $\eta$ and $\eta'$ can become large, provided they do so over a small enough range in
$\phi$. Features are thus associated with a characteristic physical scales and thus generate scale-dependent power spectra and
higher correlation functions   \cite{Chen:2006xj}.

The 3-point correlation function 
$\langle\zeta(\bk_1)\zeta(\bk_2)\zeta(\bk_3)\rangle$
is dominated by the $\epsilon \eta'$ term \cite{Chen:2006xj},
\begin{equation}
i \left( \prod_i u_{i}(\tau_{end}) \right)
\int_{-\infty}^{\tau_{end}} d\tau a^2 \epsilon  \eta' 
\left( u_{1}^*(\tau) u_{2}^*(\tau) \frac{d}{d\tau} u_{3}^*(\tau)
+ {\rm sym} \right) (2\pi)^3 \delta^3(\sum_i \bk_i) + {\rm c.c.}
~.
\label{eqn:term4}
\end{equation}
In \cite{Chen:2006xj}, we introduced the $\CG$ to describe   non-Gaussianities  with both shape and scale dependence:
\begin{equation}
\frac{\CG(k_1,k_2,k_3)}{k_1 k_2 k_3}
\equiv \frac{1}{\delta^3(\bk_1+\bk_2+\bk_3)}\frac{(k_1
k_2k_3)^2}{\tilde{P}^2(2\pi)^7}
\langle\zeta(\bk_1)\zeta(\bk_2)\zeta(\bk_3)\rangle  
\label{eqn:g},
\end{equation}
where ${\tilde P}$ is a constant, and for convenience we set it to be   roughly equal to  the magnitude of the power spectrum,  $\tilde{P}^2 \equiv
4\times 10^{-10}$. In the absence of the sharp feature,  (\ref{eqn:g}) reduces to the local form with $\CG = (3/10)
f_{NL}^{local} \sum k_i^3$ in WMAP's convention.  We now construct an analytic approximation to this function.

\begin{figure}[t]
\myfigure{5in}{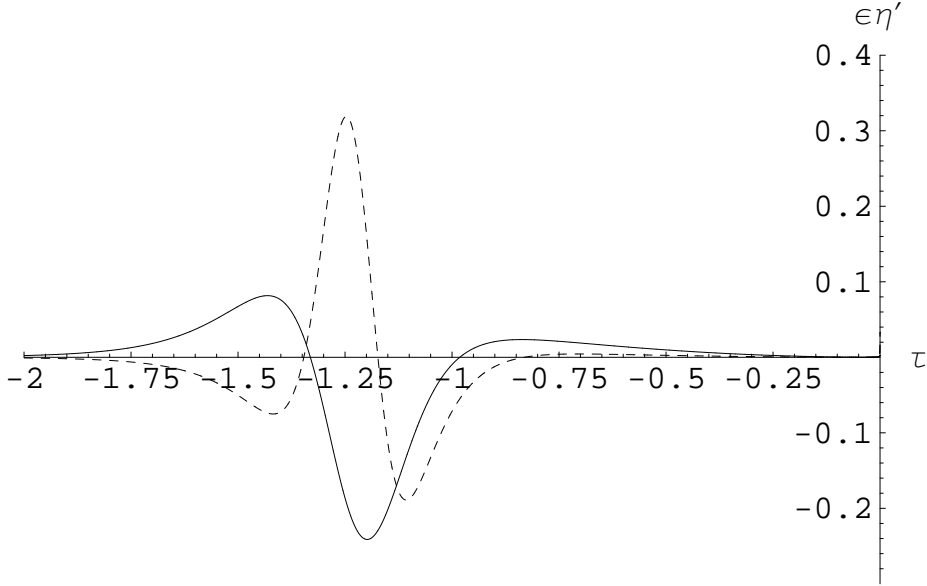}
\caption{The evolution of $\epsilon\times \eta'$ with units $k_*\tau =
-1.2$ where $k_*=1$ is set to be the scale when $\phi$ crosses the
center of the feature for our two models . The step ($c=0.0018, d=0.022M_p, \phi_s = 14.84M_p$) is the solid line, and bump is the dashed line   $(c=0.0005,
d=0.01M_p, \phi_b
= 14.84M_p)$. In both cases $\eta'$ is boosted by  $\CO(1000)$ boost
for around one Hubble time ($\delta \tau \approx 1$ in our
units). } 
\label{fig:epsilonetadottotal}
\end{figure}

\begin{figure}[t]
\myfigure{3in}{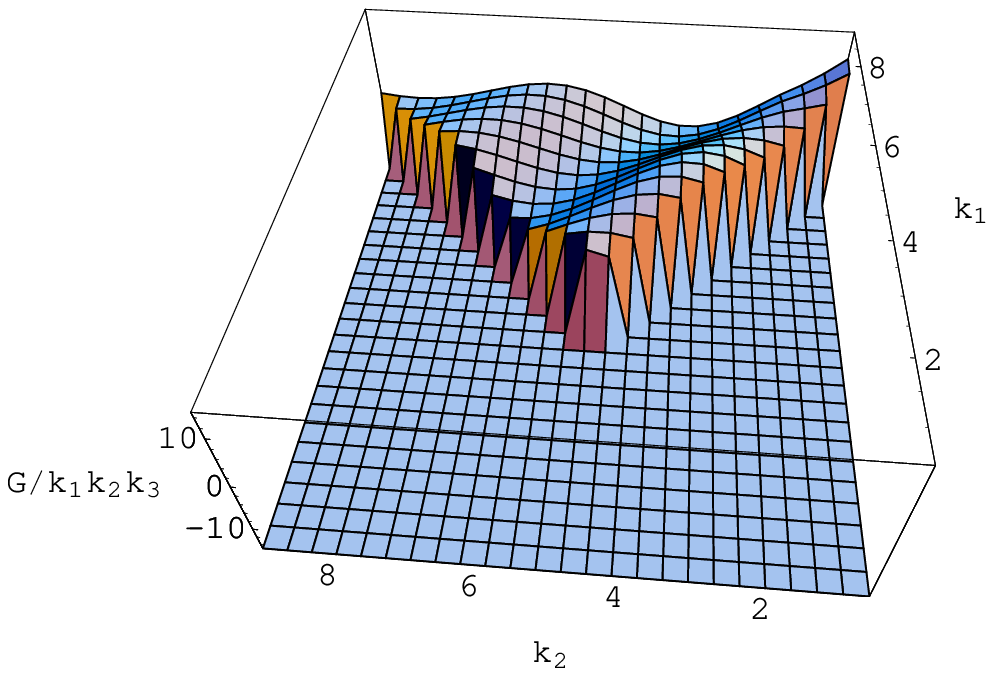}
\myfigure{3in}{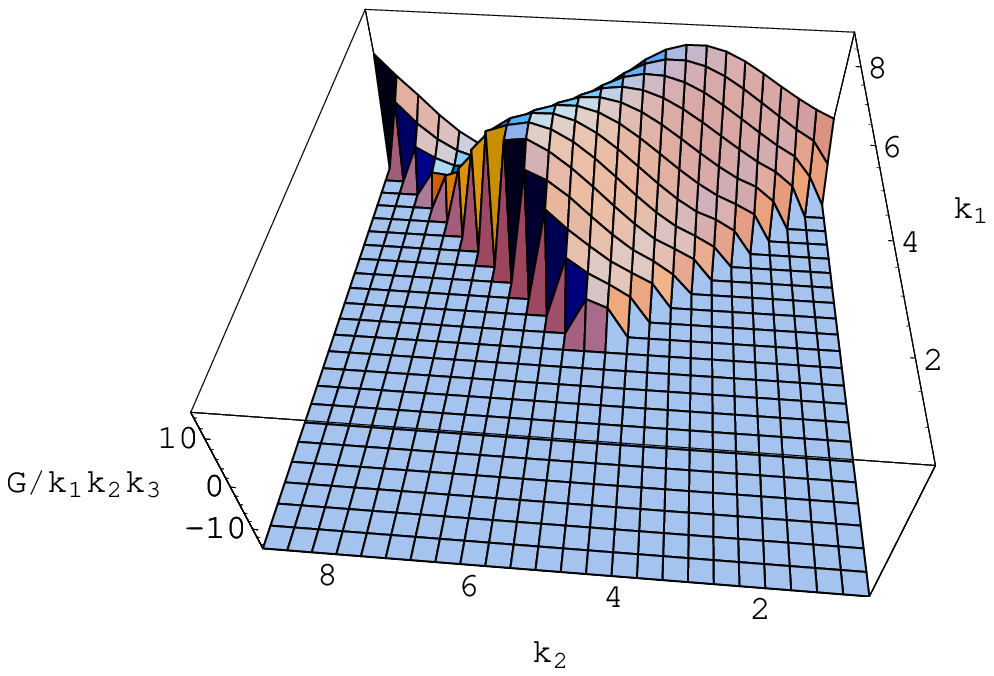}
\caption{A sample result of the step (left) and bump (right) models,
where for both plots $k_3=9$. We numerically computed the 3-point
correlation functions of both models for $1 < k < 9$ such that $k=1/1.2$
corresponds to the scale of the feature at $\phi_s = 14.84M_p$. For the
step model, we use $(c=0.0018, d=0.022)$ while for the bump model we use
$(c=0.0005, d=0.01)$.}  
\label{fig:feature3dplot} 
\end{figure}

To illustrate our approach, we consider two specific features, the step \cite{Adams:2001vc, Chen:2006xj} 
\begin{equation}
V(\phi)=\frac{1}{2}m^2\phi^2\left[1+c \tanh\left(\frac{\phi-\phi_s}{d}\right)\right], \label{eqn:potentialstep}
\end{equation}
and the bump  
\begin{equation}
V(\phi) = \frac{1}{2}m^2\phi^2\left[1 + c
~\mathrm{sech}\left(\frac{\phi-\phi_b}{d}\right)\right] ~,
\label{bump}
\end{equation}
where $c$, $d$ and $\phi_b$ again respectively  determines the height, width
and location of the feature. In the latter case,   $c$ must be small enough to ensure that the field point does not get trapped in a local minimum.   We present the numerical results
for $\epsilon \eta'$ in Fig.~\ref{fig:epsilonetadottotal} and the
non-Gaussianity profile in Figure \ref{fig:feature3dplot}.

As we see from Fig.~\ref{fig:epsilonetadottotal}, $\eta'$ is non-trivial over a small range of   $\tau$,  but we do not have an analytic result for $u_i(\tau)$, and in \cite{Chen:2006xj} we performed these integrals numerically.  Now consider  a series of hat functions,
$\eta' = \eta'_m \equiv {\rm constant}$, for $\tau_m-\delta \tau_m 
< \tau <
\tau_m+\delta \tau_m$, and $0$ otherwise. Outside the range of these
hat functions, $\eta =\CO(\epsilon) \approx 0$, so 
\bea
\int d\tau \eta' = 2\sum_m \eta'_m \delta \tau_m \approx 0 ~.
\label{eta'cond}
\eea
Namely, the area under the hat functions or the numerical curve of $\eta'$ 
should sum to approximately zero.
We approximate  $\zeta(k_i)$ by its unperturbed form
\begin{equation}
\zeta(k_i) = \frac{i H}{\sqrt{4\epsilon k_i^3}} (1+i k_i
\tau)e^{-i k_i \tau} ~,
\end{equation}
although the actual mode functions can deviate from
this simple form by as much as a factor of two (in addition to phase
shifts) for short periods of time \cite{Adams:2001vc}.
Using these approximations,
\bea
\langle \zeta(\bk_1)\zeta(\bk_2)\zeta(\bk_3) \rangle 
&\approx&
i\frac{H^4}{64 \epsilon^3 \prod_i k_i^3}
(2\pi)^3 \delta^3 (\sum_i \bk_i) \nonumber \\ &&
\times \sum_m \epsilon \eta'_m
\int_{\tau_m-\delta \tau_m}^{\tau_m+\delta \tau_m}
\frac{d\tau}{\tau} 
\left(1-i(k_1+k_2)\tau - k_1k_2\tau^2 \right) k_3^2 e^{i K\tau}\nonumber \\ &&
+ {\rm sym} + {\rm c.c.}  
\label{Integration} \\
&=& (2\pi)^7 \delta^3(\sum_i \bk_i) P_k^2 \frac{1}{\prod_i k_i^3}
\times
\sum_m \frac{\eta'_m}{8}
\left[ -\sum_i k_i^2 ~{\rm Im} {\rm Ei}(i K\tau) \right. \nonumber \\ 
&& \left.+ \frac{\sum_{i\ne j} k_i k_j^2}{K} \sin K\tau
- \frac{k_1k_2k_3}{K} (\sin K\tau -K\tau\cos K\tau)
\right]_{\tau_m-\delta \tau_m}^{\tau_m+\delta \tau_m} ~,
\nonumber \\
\label{ansatzderi}
\eea
where 
\bea
K\equiv k_1 + k_2 + k_3 ~,
\eea
and ${\rm Im Ei}(i K\tau)$ denotes the imaginary part of the exponential
integral function.
For $K\tau\gg 1$, ${\rm
Im Ei}(i K\tau) \to -\pi-\cos(i K\tau)/(K\tau) + \CO((K\tau)^{-2})$; for $K\tau\ll 1$, 
${\rm Im Ei}(i K\tau) \to K\tau + \CO((K\tau)^2)$.

\begin{figure}[t]
\myfigure{3in}{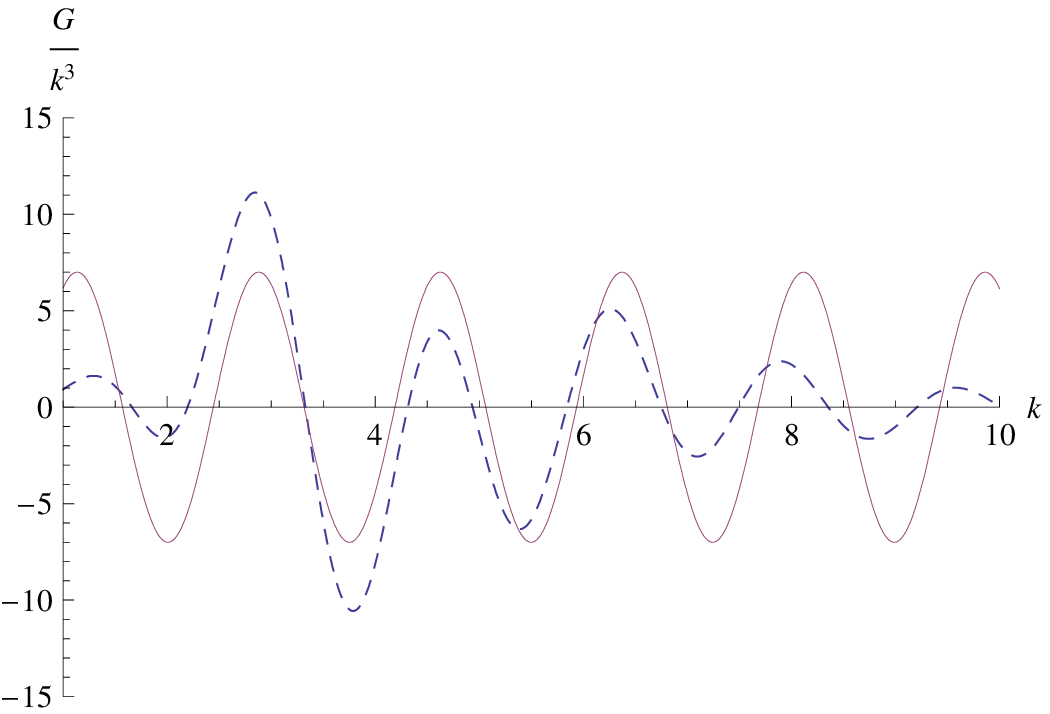}
\myfigure{3in}{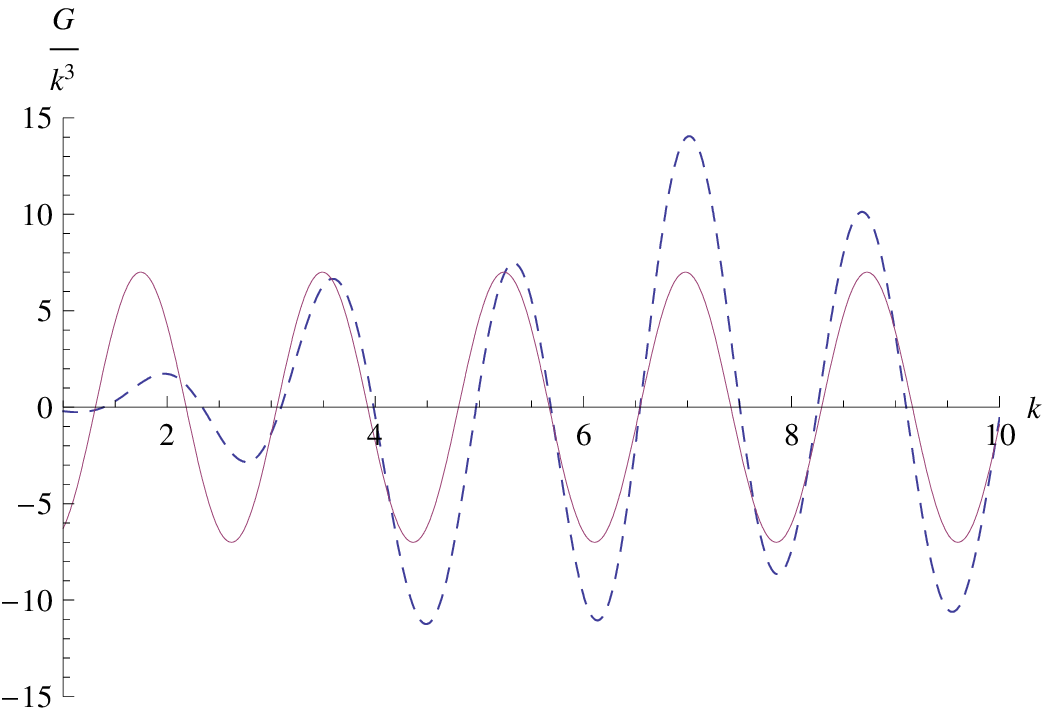}
\caption{A comparison of the ansatz Eq. (\ref{SFansatz}) (full
line) to our numerical results (dashed line) for the step model $(c=0.0018,d=0.022)$ on the left and the bump model $(c=0.0005,d=0.01)$ on the right with $k_*\approx 1/1.2$. This is a reasonable match to our analytical from. As explained in the text, the drop-off at small and large $k$ is not captured by the ansatz but can be easily incorporated by adding in a window function when comparing the ansatzen with data. We have added a phase factor into the ansatz to synchronize with
the numerical results; this phase factor is physically important but is
not estimated analytically.}
\label{fig:approxvsnumerical}
\end{figure}

For large $K$, $K \tau_m \gg 1$, Eq.~(\ref{ansatzderi}) is
dominated by the last term,
\bea
\frac{\CG}{k_1k_2k_3} \sim 
-\frac{1}{4} \sum_m \eta'_m \tau_m \sin K\tau_m \sin K\delta \tau_m ~.
\label{ansatzlargek}
\eea
Numerically, the dominant scale-dependent oscillation comes from the term $\sin K\tau_m$, where
$\tau_m$ is the center of the hat function, and corresponds to
the moment the inflaton crosses the feature and $\phi(\tau_m) \approx
\phi_s$. We denote this scale  $-1/k_*$. 
The amplitude  is further modulated by $\sin
K\delta \tau_m$. Typically $\delta\tau_m \ll \tau_m$, so this
terms is less important unless the non-Gaussianities are found over a large $k$-range. 
From (\ref{eta'cond}),  we see the
amplitude of (\ref{ansatzlargek}) is roughly $\eta' \Delta \tau$
($\Delta \tau$ is the duration of the feature in $\eta'$). This amplitude is consistent with the
order-of-magnitude estimate of
Ref.~\cite{Chen:2006xj}.

Before we write down the ansatz, let us take a step back and consider  
several other limits not well described by
Eq.~(\ref{ansatzlargek}).
For small $K$, $K\tau_m \ll 1$, and the leading terms (the first two terms in (\ref{ansatzderi})) are proportional to
$\sum_m \eta'_m \delta \tau_m \approx 0$. So the next order terms
dominate,
\bea
\frac{\CG}{k_1k_2k_3} \sim \sum_m \eta'_m \delta \tau_m \tau_m^2 K^2 ~.
\label{ansatzsmallk}
\eea
The non-Gaussianities vanish as $K\tau_m \to 0$ as these modes are
outside the horizon as the inflaton encounters the feature.  

 In the squeezed triangle limit, $k_1=k_2=k$ and $k_3\to 0$, the second
term in (\ref{ansatzderi}) dominates instead of the third and we have a different behavior
\bea
\frac{\CG}{k_1k_2k_3} \sim \sum_m \frac{\eta'_m}{4 k_3} \cos 2k\tau_m
\sin 2k \delta \tau_m ~.
\label{ansatzsq}
\eea
Comparing (\ref{ansatzlargek}) and (\ref{ansatzsq}), we see that the
latter becomes more important if one of the momenta $k_3$ is smaller
than $1/|\tau_m|$. This is less important observationally than Eq. (\ref{ansatzlargek}),
as it requires measurements on two widely different scales.
In addition, Eq.~(\ref{ansatzlargek}) is incomplete also because it does  not
vanish for large $K$, and this is an artifact of the sharp edge of our
hat functions. We expect (\ref{ansatzlargek}) to decay away after a
width $\Delta K > 1/\delta\tau$ where $\delta \tau$ is the smoothing
of the sharp edge of the hat function. 

With these caveats in mind, we can now write down the ansatz following Eq. (\ref{ansatzlargek}) by defining
\begin{equation}
\frac{\CG_{\mathrm{feat}}}{k_1k_2k_3} \equiv f_{\rm feat} 
\sin \left(\frac{K}{k_*}+\mathrm{phase}\right) ~,
\label{SFansatz}
\end{equation}
where we match the phase to our numerical results.  We could improve this approach by  taking a sum of a series of hat functions, 
so the final form of the
3-point correlation function would then be a sum over oscillatory
functions
\begin{equation}
\frac{\CG_{\mathrm{feat}}}{k_1k_2k_3} =\sum_m  f_{\rm feat}^{(m)}\sin
\left(\frac{K}{k_*^{(m)}}+\mathrm{phase}\right) ~,
\label{eqn:sumansatz}
\end{equation}
where $k_*^{(m)}$ is the scale associated with the center of the
$m$-th hat function, or further polish this result using an Euler-Maclaurin expansion.  However, our primary goal here was to produce  an analytic expression that mirrored the qualitative form of the 3-point function generated by a step, and in this we have succeeded.     This expression is scale-dependent and thus differs from ``local'' and ``equilateral'' forms. Finally, this ansatz is clearly factorizable and in future work we will use it -- in combination with the algorithms of \cite{Smith:2006ud} -- to probe the ability of specific CMB experiments to recover this type of signal from data.

For the step potential \cite{Chen:2006xj},
\bea
f_{\rm feat} \sim \frac{7c^{3/2}}{d\epsilon} ~,
\label{fStepEst}
\eea
where $c$ and $d$ are the height and width of the step respectively.
This feature is approximately localized, with a single scale $k_*$.
The derivation of the 
phase factor is not instructive and we obtained it by matching to our
numerical results.
Refs.~\cite{Peiris:2003ff,Covi:2006ci} use a step to explain the
``glitch'' seen in the temperature $C_\ell$ for $\ell \sim 30$, and
find best fit parameters $(c=0.0018,d=0.022,\phi_s =
14.84\mpl)$. Looking at
Fig.~\ref{fig:approxvsnumerical} we find 
\begin{equation}
\frac{\CG_{\mathrm{feat}}}{k_1k_2k_3}   = 7 \sin
\left(\frac{K}{0.83}+\mathrm{phase}\right) ~.
\end{equation}
 The amplitude is within the order of magnitude
that we expected from Eq.~(\ref{fStepEst}) and $\tau_m =
-1/k_* \approx -1.2$ is indeed around the center of the feature in
Fig.~\ref{fig:epsilonetadottotal}.   These approximations are intended to capture the qualitative form of the 3-point function and do not have a high degree of numerical fidelity, but could certainly be further developed. Moreover, the detailed numerical match to the exact result could  be improved substantially by adding heuristic parameters to the approximation, and then varying these to optimize the   approximation. 

\subsection{Non-attractor initial conditions}
\label{Sec:IC}

Non-Gaussianities can also be enhanced if the inflaton trajectory is initially displaced from its attractor, slow-roll solution -- for instance by giving it large kinetic energy (``fast-roll'') or perturbing the field point in a steep direction, orthogonal to the inflaton trajectory. 
Hubble friction will erase this transient, so this situation  only arises when the overall duration of inflation is close to the minimal value. Again, modes that are crossing the horizon during this period
have their non-Gaussianities boosted due to the enhanced coupling. 
In this case the relevant scales are the longest modes that contribute to the CMB.    

This mechanism was used by 
Contaldi et.~al.~\cite{Contaldi:2003zv} to explain the
possible low-$l$ suppression in CMB (see also \cite{Boyanovsky:2006qi,Boyanovsky:2006pm}).\footnote{Ref.~\cite{Powell:2006yg} extended this approach to other
non-inflationary initial conditions, and the same conclusion applies to this model. Conversely, \cite{Burgess:2002ub} looks at a perturbation in a two-field model, which is beyond the scope of the current analysis.}  The inflaton is started with a large
velocity and the potential is negligible, so 
\bea
\ddot \phi + 3H \dot \phi \approx 0 ~, ~~~~ 
H^2 \approx \frac{\dot \phi^2}{ 6 } ~.
\eea
One finds  $a(t) \sim t^{1/3}$, and  $\dot \phi
\approx - \sqrt{6}/(3(t+t_0))$, where $t_0$ is determined by the
initial velocity. 
During this fast-roll period,
\bea
\epsilon_f \approx 3 ~, ~~~~~ \eta_f \approx 0 ~.
\label{epsilon_f}
\eea

In a qualitative estimate,
Ref.~\cite{Contaldi:2003zv} assumes that the slow-roll inflation
period begins immediately after this kinetic-energy-dominated
fast-roll (non-inflationary) period, and during slow-roll
\bea
\epsilon_s \approx \eta_s \approx 0.01 ~.
\eea

Let us estimate the non-Gaussianity. For example, consider (\ref{eqn:term5}) 
\begin{equation}
\frac{i}{2} \int_{\tau_{\rm begin}}^{\tau_{\rm end}} 
d\tau~ a^2 \epsilon^3
\left(\prod_i u_{i}(\tau_{end}) \right) 
\left( u^*_{1} \frac{du^*_{2}}{d\tau} \frac{du^*_{3}}{d\tau}
\frac{\bk_1\cdot \bk_2}{k_2^2} + {\rm five~perm} \right)
(2\pi)^3 \delta^3(\sum_i \bk_i) + {\rm c.c.} ~.
\label{term5}
\end{equation}
The fast-roll period is not inflationary, so modes are entering the
horizon during this phase. As we will explain more quantitatively in
Appendix \ref{App:IC}, we are interested in the modes that are slightly
within the horizon as inflation begins. Modes that are near the
horizon have smaller non-Gaussianities.
Although the coupling $\epsilon_f
\approx 3$ is greatly enhanced comparing to $\epsilon_s \approx 
0.01$ in the slow-roll
phase, the mode function $u_k \propto 1/\sqrt{\epsilon_f}$ is greatly
suppressed. Looking at  (\ref{term5}), we find the following  factors of $\epsilon$ in the amplitude of the 3-point correlation  
\bea
\frac{\CG}{k^3} \sim f_{\rm fast} \sim \epsilon_f^3 \epsilon_s^{-3/2} \epsilon_f^{-3/2}
\epsilon_s^2 ~.
\label{fNLfast}
\eea
The first factor is the large coupling; the second is
due to the asymptotic value $u_k(\tau_{end})$, which is determined
by the slow-roll inflation period; the third is from
$u_k(\tau)$ during the fast-roll period -- 
note this is where the suppression
comes from; the fourth is from the prefactor, $1/\tilde P^2 \propto
\epsilon_s^2$, in the definition of $\CG/k^3$. 
This gives
\bea
f_{\rm fast} \sim 0.5 ~,
\eea
which is confirmed by a more detailed estimate in Appendix
\ref{App:IC}.

We next look at the transition period from the end of the fast-roll to
the slow-roll, where $\epsilon$ drops from $\epsilon_f \approx 3$ to
$\epsilon_s \approx 0.01$.  
This period is absent in the above
analytical model of Ref.~\cite{Contaldi:2003zv}, but it also has
important contribution to the non-Gaussianities.
Once the kinetic and potential energies are roughly balanced,  $H$ is approximately constant, and  $\dot \phi$ drops as
$e^{-3Ht}$ and $\epsilon$ drops as $e^{-6Ht}$. The $\eta$ grows from
$0$ to $\CO(1)$ at the beginning of this period, then drops to
$\CO(0.01)$ in the end. The most important contribution to the
3-point is the $\epsilon \dot \eta$ term, similarly to the
sharp feature case, and possess the amplitude
\bea
f_{tran} = \CO( \Delta \eta) = \CO(1) ~.
\eea
The shape and running of this non-Gaussianity is also similar to those
of the sharp feature, hence the ansatz (\ref{SFansatz}) applies.

The sum of these two contributions implies that the 3-point is 
$\CO(100)$  larger than the $f \approx \epsilon_s$ of
standard slow-roll inflation. Our numerical simulation (not shown here) confirms this expectation. This is only comparable to the
$f_{NL}^{local}=\CO(1)$  3-point expected from simple
non-linear gravitational effects of \cite{Pyne:1995bs}, and an order of magnitude smaller than the 3-point we found for a bump.  Since the kinetic energy is simply redshifted away by the expansion of the universe, we cannot easily enhance this signal by tuning the potential.  Further, this signal peaks at scales where cosmic variance will be largest,  so we are not optimistic this signal will be observable, even under ideal conditions.

\section{Sub-horizon generation of non-Gaussianities: Resonance
model} 
\label{sect:subhorizon} 

We now turn our attention to the generation of significant
non-Gaussianities while the modes are still well within the
horizon. Again, we inspect the integral representation of the 3-point
correlation function, Eq. (\ref{eqn:Itype1}).  Modes that are well within the horizon oscillate rapidly. Since the interaction couplings
$\epsilon,\eta'$ are roughly constant for plain vanilla slow-roll
model, these oscillations cancel. However, if the interaction
couplings  oscillate, they can interfere \emph{constructively} with the
rest of the integrand, yielding an enhanced 3-point signal. These contributions are generated while the modes are deep inside the  horizon, and are thus physically distinct from the situation explored previously.  

\begin{figure}[t]
\myfigure{5in}{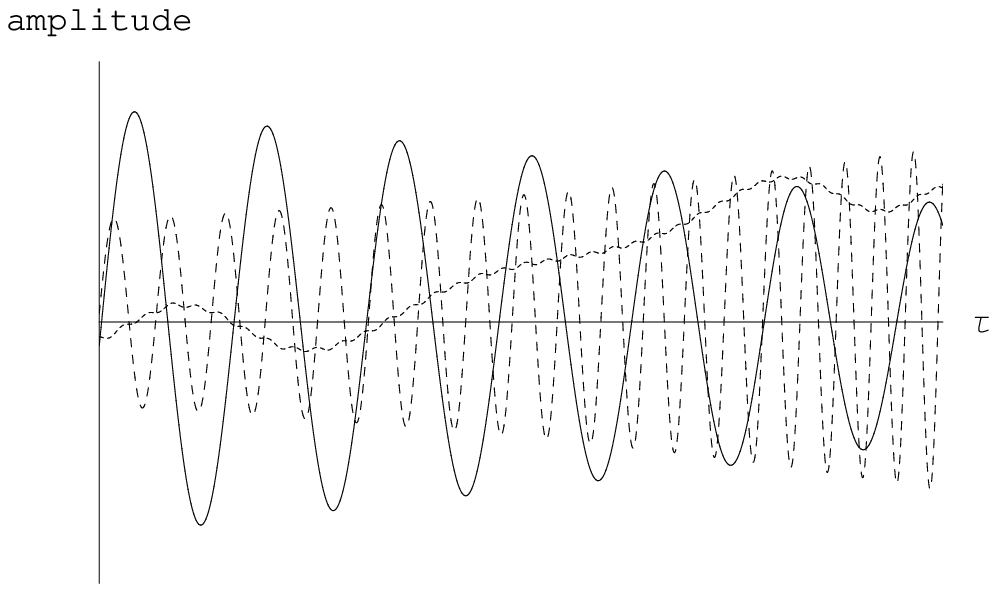}
\caption{This figure illustrates resonance between the total mode
$K=k_1+k_2+k_3$ momentum and an oscillatory $\eta'$.  The solid line shows $\zeta(k)$, the dashed line describes $\eta'$, and the dotted line is the integral
Eq.~(\ref{eqn:Itype1}) \emph{from $-\infty$ to time $\tau$} with $k_1=k_2=k_3=k$. Resonance occurs when the frequency of $\eta'$ is roughly $3k$.   This figure was generated after numerically evaluating the mode functions for the parameters of  Eq. (\ref{Nparameters}), with all the lines rescaled to arbitrary units to emphasise the effect. The universe grows by roughly one e-fold over the range of this plot, and the relevant modes are well inside the horizon.} 
\label{fig:resonanceevol}
\end{figure}

When all modes are inside the horizon, each of the 3-point integrals
(\ref{eqn:Itype1}) consist of  an oscillatory piece $\sim e^{i(k_1 +
  k_2 + k_3) \tau}$ and a prefactor $g(\epsilon,\eta')$ which is a
function of the slow roll parameters.  If the potential has a small
oscillatory component, $g$ becomes
\begin{equation}
g(\epsilon,\eta') \sim
\alpha  (1 + \delta \sin (\omega \tau)) ~,
\end{equation}
where $\omega > H$, and $\delta \ll 1$.  We assume that $\alpha$, $\omega$ and $\delta$ change slowly over a single Hubble time, and treat them as constants in what follows. As the physical wavelength
$a(t)/K$ increases, the mode 
will briefly resonate with the coupling term $g$ when its frequency is
roughly $\omega$. During this resonance phase, the $\xi_1 \xi_2 \xi_3$
term in the integrand of (\ref{eqn:Itype1}) can interfere
constructively    $\sin (\omega
\tau)$ and generate a large 3-point
function. Conversely, destructive interference will generate no extra contribution to the 3-point function. See Fig.~\ref{fig:resonanceevol} for an illustration of the resonance effect.

This resonance requires  
\bea
H < \omega < \mpl ~,
\label{ResCond}
\eea
in order to ensure that the relevant modes are sub-Planckian during
the resonance epoch. In practice, this is not a strong constraint, but
the resonance introduces a new length-scale into our analysis of
cosmological perturbations.\footnote{In models with a non-trivial UV
  cutoff below the Planck scale, this scale would replace $\mpl$ in
  (\ref{ResCond}).}  In this case, we expect non-Gaussianities to be
present at all scales, since the ripple is laid down across the entire
potential,  in contrast to  the ``features'' considered above.
Moreover, the modulation of the potential must be small enough to
ensure that the inflaton does not get trapped -- and if we make it
small enough, we can also ensure that the 2-point function is not
significantly modified even if the non-Gaussianities are large, as
discussed in Appendix \ref{App:Res}.

To study this mechanism explicitly, consider a standard slow-roll model with a very small oscillatory component
\begin{equation}
V(\phi) = 
\frac{1}{2}m^2 \phi^2 \left[1 +
c~\sin\left(\frac{\phi}{\Lambda}\right)\right] ~.
\label{Vexample}
\end{equation}
Since $\phi$ is slowly rolling, the resonance effect occurs when
the physical frequency reaches
\begin{equation}
\omega  \approx \frac{\dot\phi}{2\pi \Lambda} =
\frac{m}{\sqrt{6}\pi \Lambda} ~.
\end{equation}
Therefore to satisfy the resonance condition (\ref{ResCond}),  we need
$\Lambda \phi \ll 1$ and $m\ll \Lambda$, while we take
$H=m\phi/\sqrt{6}$, in accordance with slow roll. In Appendix
\ref{App:Res} we show that we need $c\phi/\Lambda \ll 1$, so the
perturbation to $\eta$ is small.

Before we proceed, we emphasize that (\ref{Vexample}) is simply a toy model, with no direct physical motivation.  However, in brane inflation, it has been argued that the duality cascade within a throat will leave tiny sharp features in the inflaton potential or the background warp factor \cite{Hailu:2006uj}. Such features
typically come in a series and can have cosmologically observable
effects as the inflaton branes roll across them \cite{Bean:2008na}. 
If these features are large and well-separated, they would yield of sequence of isolated sharp features.  However, in the opposite limit, one would have a situation closer to the modulated potential above.  

\begin{figure}[t]
\myfigure{3in}{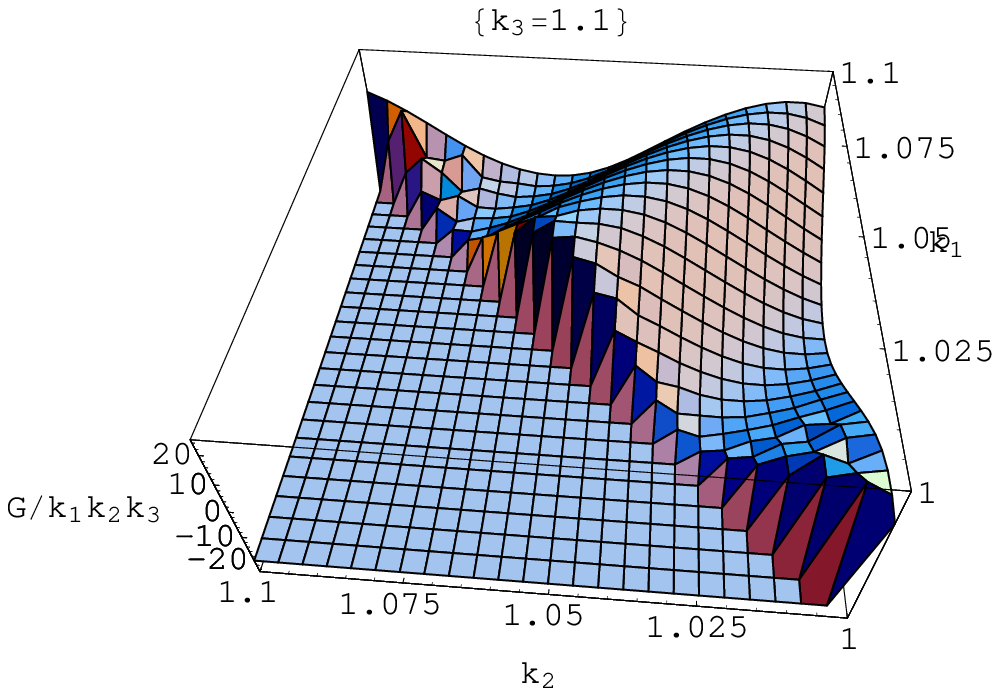}
\myfigure{3in}{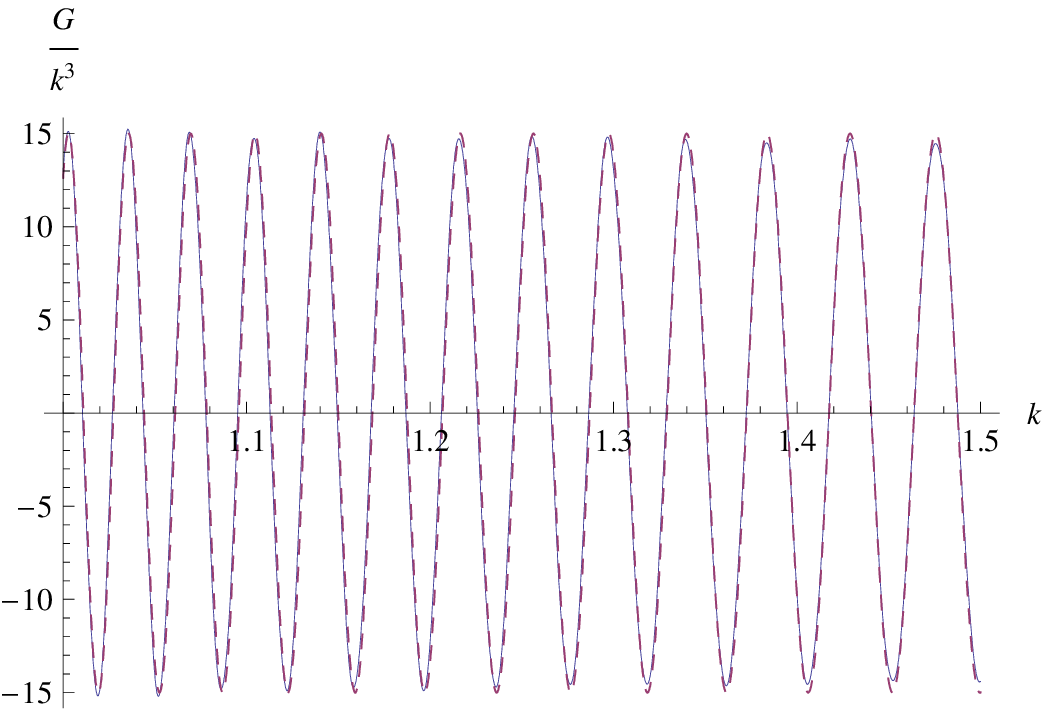}
\caption{The numerical results for the 3-point correlation function of
the resonance model. The plot on the left hand side is a 2-D slice of
the 3-D function with fixed $k_3=1.1$ while $k_2$ and $k_3$ runs from
$1.0$ to $1.1$. The plot on the right hand side is a comparison of the numerical result (full line) and our analytical ansatz (dashed line) for the
equilateral case with $1.0 < k < 1.5$ -- we have increased the range
to more  show the $1/K$ dependence of the frequency. The amplitude
(\ref{ResAmp}) over-predicts by about 30\% -- here we are have used
instead numerically computed value of $15$ to fit the plot better. On
the other hand, the frequency (\ref{ResC}) is accurate -- we have used $C=
2.05/(\phi \Lambda)$ and added a phase factor to synchronize
the plots.}  
\label{fig:resonanceresults}
\end{figure}

We now estimate the  resulting 3-point function. For definiteness, we use the following parameters in our numerical examples
\bea
m=3\times 10^{-6} M_p ~~, ~~ c=5\times 10^{-7} ~~,
~~ \Lambda=0.0007M_p ~~, ~~\phi \approx 15 M_p ~~.
\label{Nparameters}
\eea
The non-Gaussianity oscillates with $K$ because $g(\epsilon,\eta')$ 
has a continually changing phase.  The time it takes the
inflaton to cross one  ``ripple'' is $\Delta t = \sqrt{6}
\pi \Lambda/m$, during which $\Delta N_e = \pi \phi
\Lambda$. So the oscillation period of this non-Gaussianity in
$K$-space is 
\bea
\Delta K = K \Delta N_e = \pi K \phi \Lambda ~.
\label{DeltaK}
\eea
Using the parameters in (\ref{Nparameters}), this gives $\Delta
K/K=0.033 $. 
This agrees with the numerical results shown in
Fig.~\ref{fig:resonanceresults}. 
We emphasize that the property $\Delta K \propto K$ is
model-independent, since, ignoring the slow-variation of background
evolution, the analysis for $\Delta t$ and $\Delta N_e$ is
scale-independent. Another universal property is $\Delta K/K <1$, via (\ref{ResCond}).

We now estimate the amplitude of these non-Gaussianities.
The frequency of each mode is continuously decreasing due to the
expansion of the universe. Once it differs from the resonant frequency
by $\Delta \omega$, the integration in the 3pt starts to cancel out if it
is performed over $\Delta t_1 \sim \pi/\Delta \omega$.   
Meanwhile, it takes
$\Delta t_2 \sim \Delta \omega / (\omega H)$ to stretch a mode sufficiently in order to induce a frequency change  $\omega$ to $\omega - \Delta \omega$. Equating
$\Delta t_1$ and $\Delta t_2$ gives the time period over which
resonance occurs for this particular mode,
\bea
\Delta t \sim \sqrt{\frac{\pi}{\omega H}} ~. 
\eea
In  (\ref{Vexample}), this corresponds to the number of
oscillation cycles
\bea
\frac{\Delta t}{T} \approx \frac{1}{\sqrt{\Lambda \phi}}
\label{numperiods}
\eea
we need to integrate over, where $T=\sqrt{6}\pi\Lambda/m$ is
the oscillation period of the background.
For (\ref{Nparameters}) we need about $10$ cycles, which is
confirmed numerically in Fig.~\ref{fig:resonanceevol}.

The dominant source of 3-point terms -- as for the ``feature'' models -- is the $\epsilon \eta'$ interaction term Eq.~(\ref{eqn:term4}).
Because the mode is well within the horizon $|K\tau| \gg 1$, in
(\ref{Integration}) the last term dominates. Denoting the oscillatory
behavior of the slow-roll parameter as
\bea
\dot \eta \supset \dot\eta_{\rm osci} = \dot\eta_A \sin (\omega t) ~,
\eea
where the subscript ``A'' 
denotes the oscillation amplitude. Integrating
over one period 
\bea
\int d\tau ~\tau~ \sin (\omega t)~ e^{i K\tau} 
\eea
gives the amplitude $\pi \tau_* /K$, where $\tau_* = -1/aH$ is
evaluated around the resonant point. 
Combining this with Eq.~(\ref{numperiods}), from (\ref{Integration}) 
we get the estimate of the amplitude
\bea
f_{\rm res} \approx 
\frac{3}{8}\frac{ \dot \eta_A } {H \sqrt{\Lambda\phi}} ~.
\label{Gamplitude}
\eea

To evaluate (\ref{Gamplitude}), 
we need $\dot{\eta}$.
For (\ref{Vexample}) the amplitude of the dominant oscillating term 
in $\dot\eta$ is (see Appendix~\ref{App:Res})
\bea
\dot \eta_A \approx \sqrt{6}\frac{cm\phi}{\Lambda^2} ~.
\label{dotetaA}
\eea
Note one might naively expect
that $\dot{\eta} \sim V''' \sim c/\Lambda^3$, i.e. that it scales as
$\Lambda^{-3}$. However this over predicts the amplitude by a factor of $\Lambda^{-1}$ : although $\dot{\eta}
\propto \dddot{H}$, since $\dot{H} = -\dot{\phi}^2/2$ via the
scalar field equation of motion (\ref{eqn:scalareom}) and hence
$\ddot{H} = -\dot\phi\ddot{\phi} = 3H\dot{\phi}^2+V'\dot{\phi} \propto
V'$. It follows that $\dot{\eta}\propto \dddot{H}\propto V''$. The
crucial point is that one can always use Eq. (\ref{eqn:scalareom}) to
eliminate one $\phi$ derivative of $V$ in the derivation of $\dot{H}$;
in other words one cannot simply neglect the $\ddot{\phi}$ term in the
equation of motion. Numerical
computations of $\dot{\eta}$ show that this estimate
Eq. (\ref{dotetaA}) is accurate in both amplitude and scaling.

\begin{figure}[t]
\myfigure{5in}{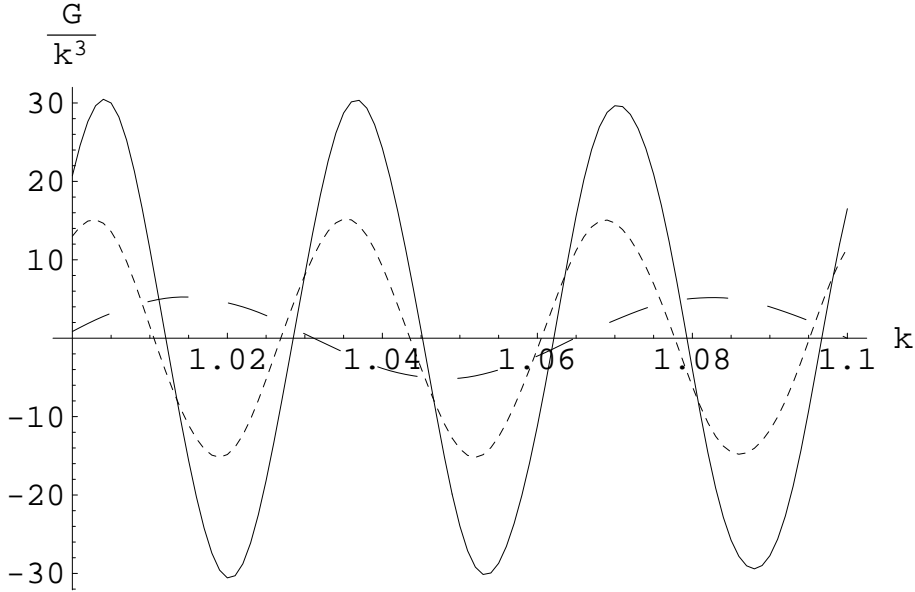}
\caption{The numerical results for the set of parameters $(c=5\times
10^{-7}, \Lambda=0.0007M_p, \mathrm{shortdashed})$, $(c=10 \times
10^{-7}, \Lambda=0.0007M_p, \mathrm{full})$ and $(c=10\times 10
^{-7}, \Lambda=0.0014M_p, \mathrm{longdashed})$. The amplitude of the 3-point
correlation function is proportional to $c/\Lambda^{2.5}$ 
while the frequency is independent of
$c$ and proportional to $1/\Lambda$.} 
\label{fig:resonanceparams}
\end{figure}

Combining (\ref{Gamplitude}) and  (\ref{dotetaA}), we get
the amplitude 
\bea
f_{\rm res} 
&\approx& \frac{9}{4}\frac{c M_p^3}{\Lambda^{2.5}\phi^{0.5}}
~,
\label{ResAmp}
\eea
where we have restored $M_p$ into the final answer.
For (\ref{Nparameters}),  $f_{\rm res} \sim 22$, which differs from matches to  
our exact numerical results shown in
Fig.~\ref{fig:resonanceresults} by $\sim30\%$. The numerical results also exhibit
the $\CG/k^3 \propto c/\Lambda^{2.5}$ scaling, as can been seen in
Fig.~\ref{fig:resonanceparams}.

The ansatz for the non-Gaussianity is therefore
\bea
{\frac{\CG}{k_1k_2k_3}}_{\rm ~ansatz}
= f_{\rm res} \sin( C \ln K+ \mathrm{phase}) ~,
\label{eqn:resonanceansatz}
\eea
where
\bea
C =2\pi K/\Delta K \approx 2/(\phi\Lambda) ~.
\label{ResC}
\eea
The $\ln K$ comes from the fact [in Eq.~(\ref{DeltaK})] that the
oscillation period $\Delta K$
is proportional to $K$ and the background $\phi$ dependence of $\Delta
K/K$ is approximately scale-invariant.
The value of $C$
is determined by the oscillation frequency in the $K$-space. 
We also point out that the power spectrum has the same oscillatory
frequency in $K$-space, due to the same reason that we stated before
(\ref{DeltaK}). See Fig.~\ref{fig:resonancepower} in
Appendix \ref{App:Res} for details. We compare this ansatz with the
numerical result in Fig.~\ref{fig:resonanceresults}.
This ansatz is factorizable at least by Taylor-expanding $\ln K$ in a range smaller than $K$. Note unlike the ansatz for the feature (\ref{SFansatz}) where
 the non-Gaussianities are peaked around some fixed scale
 $k_*$,  the amplitude does not decay and is present at all scales. In
 addition, $K > \Delta K$  so it oscillates faster.  As we noted
 above, the specific potential we consider is a toy model, but this
 analysis could be extended to more complicated modulations, or models
 where the modulation applied only over a finite range of field
 values.

\section{Discussion and Conclusion} \label{sect:Conclusions}

We studied two distinctly mechanisms for
generating large non-Gaussianities  within 
single field inflation. We derive the approximate 3-point
correlation functions using semi-analytic methods, which
are in the computationally useful \emph{factorizable} form.  In the
first mechanism, non-Gaussianity is generated at horizon crossing by
either a feature in the potential, or an initial transient in the
inflationary dynamics.  In the second case, oscillating slow roll
parameters induce a resonance which leads to the generation of a
non-Gaussian signal well before horizon crossing. In both cases,
the 3-point function depends strongly on the individual wavelengths of
the modes in the ``triangle''.   With a single, sharp feature or
non-standard initial conditions, the 3-point is only enhanced in modes
which are crossing the horizon as the inflaton traverses the
``feature''.   In a resonance model such as (\ref{Vexample}),  the
physical non-Gaussianity is present at all scales.

In the resonance case, we showed that the 3-point function is periodic
with a period $\Delta K$  proportional to, and smaller than, $K=k_1+
k_2+ k_3$.  This non-Gaussianity will peak starting from a fixed scale
when projected onto the CMB sky.    
Assume for
simplicity that $K \sim \ell$ where $\ell$ is the CMB multipole, and
denote $\Delta \ell$ as the oscillation period.
At larger scales where the
oscillation spanning $\Delta \ell <1$, this non-Gaussianity presumably
cannot be resolved experimentally. For the numerical example
considered here, it would become  visible at $\ell \sim
\CO(100)$ where $\Delta \ell$ starts to exceed $\CO(1)$, inducing an effective scale-dependence in this signal. 

With a feature in the potential, the resulting transient violation of
 slow-roll  generically leads to an oscillatory and
 scale-dependent  3-point function.  We wrote down a heuristic and factorizable scale-dependent expression  for this signal
 and showed that it captured the qualitative properties seen in the exact numerical evaluations of the corresponding integrals. The 3-point correlation decays as we move away from  
$K \sim k_{*}$, where $k_{*}$ corresponds to the scale leaving the horizon as the inflaton traverses the feature.  We also show that while non-standard initial conditions such as the
 fast-roll model of \cite{Contaldi:2003zv} can generate large
 non-Gaussianities relative to our usual expectations for single field inflation, the amplification is not expected to lift them above the
 ``noise'' of non-linearities produced by post-inflationary
 gravitational evolution alone.  
 
In addition,  we have presented detailed description of a general
numerical method for computing the 3-point correlation functions of
primordial perturbations from canonical single scalar field
inflationary models with arbitrary potentials. We show that while the
integrals themselves are formally convergent, they need to be
regularized as the integrands are oscillatory, and show how this can be accomplished analytically, rendering the numerical integrals rapidly convergent.  

There is much further work to be done. Our immediate goal \cite{preparation} is to use our heuristic ansatzen to construct an
optimal estimator with which to search for scale-dependent
non-Gaussianities in the cosmic microwave background, and estimate the likely bounds that future missions can put on this signal.
On the theoretical front,  we plan to investigate the details of
non-Gaussianities generated by multi-field models
\cite{Seery:2005gb,Battefeld:2006sz} within the horizon, and with
non-standard kinetic terms.  Moreover, while the analytic approximations we present here capture the qualitative form of the 3-point function, they are not intended to provide a precise quantitative match to the numerically computed values, but these approximations can certainly be improved.

\section*{Acknowledgments}

We thank Eiichiro Komatsu, Henry Tye, Jiajun Xu, Hiranya Peiris, Kendrick Smith and
Sarah Shandera for valuable discussions.
XC and EAL would like to thank the Kavli Institute for Theoretical
Physics in China and the organizers of the ``String Theory and
Cosmology'' program, where part of this work
is done, for their warm hospitality.
XC is supported by the US Department of Energy under cooperative
research agreement DEFG02-05ER41360.
RE is supported in part by  the United States Department of Energy,
grant DE-FG02-92ER-40704.

\appendix

\section{3-point correlation functions} 
\label{section:3pts}

In minimally coupled single field inflation,
the cubic interaction Hamiltonian for the scalar perturbation $\zeta$ is
\cite{Maldacena:2002vr,Seery:2005wm,Chen:2006nt}
\begin{eqnarray}
H_{int}(\tau) &=&
-\int d^3x \Bigg\{ a \epsilon^2 \zeta \zeta'^2  + a\epsilon^2 \zeta (\partial \zeta)^2  - 2 \epsilon \zeta' (\partial \zeta) (\partial \chi) 
\nonumber \\
&& + \frac{a}{2} \epsilon \eta' \zeta^2 \zeta'
+ \frac{\epsilon}{2a} (\partial \zeta) (\partial \chi) 
(\partial^2 \chi)
+ \frac{\epsilon}{4a} (\partial^2 \zeta) (\partial \chi)^2
\Bigg\} ~,
\label{Hint3}
\end{eqnarray}
where
\begin{eqnarray}
\chi &=& a^2 \epsilon \partial^{-2} \dot \zeta ~.
\end{eqnarray}
Here $\partial^{-2}$ is the inverse Laplacian. This cubic Hamiltonian is
exact for arbitrary $\epsilon$ and $\eta$.

The 3-point correlation function at some time $\tau$ after horizon
exit is 
\begin{equation}
\langle\zeta(\tau,\textbf{k}_1)\zeta(\tau,\textbf{k}_2)\zeta(\tau,\textbf{k}_3)\rangle=
-i\int_{\tau_0}^{\tau} d\tau' ~ a ~ \langle
[
\zeta(\tau,\textbf{k}_1)\zeta(\tau,\textbf{k}_2)\zeta(\tau,\textbf{k}_3),{H}_{int}(\tau')]
\rangle ~, \label{eqn:interaction}
\end{equation}
together with a term coming from the field redefinition
\begin{eqnarray}
\langle\zeta(\bk_1)\zeta(\bk_2)\zeta(\bk_3)\rangle
&=&\langle\zeta_n(\bk_1)\zeta_n(\bk_2)\zeta_n(\bk_3)\rangle
\nonumber \\
&+&\eta
\langle \zeta_n^2(\bk_1) \zeta_n(\bk_2) \zeta_n(\bk_3) \rangle
+ {\rm sym} + \CO(\eta^2 (P^\zeta_k)^3) ~,
\label{FieldRedef}
\end{eqnarray}
where $\zeta_n^2(\bk)$ denotes the Fourier transform of
$\zeta_n^2(\bx)$.

We evaluate it using the decomposition
\begin{eqnarray}
\zeta(\tau,\textbf{k})=u(\tau,\textbf{k})a(\textbf{k})
+u^*(\tau,-\textbf{k})a^{\dagger}(-\textbf{k}) ~, \\
v_k\equiv z u_k ~, ~~~~z \equiv a \sqrt{2\epsilon} ~,
\label{vdef}
\end{eqnarray}
where 
$v(\tau,\bk)$ is the solution of the linear equation of motion of the
quadratic action,
\begin{equation}
v_k'' + k^2 v_k - \frac{z''}{z} v_k =0 ~.
\label{quadeom}
\end{equation}

Our choice of vacuum implies that the initial condition for the mode function is given by the Bunch-Davies vacuum
\begin{eqnarray}
v_k(\tau_0)&=& \sqrt{\frac{1}{2k}}  \nonumber \\
v_k'(\tau_0)&=& -i\sqrt{\frac{k}{2}} \label{eqn:initialconditions}
\end{eqnarray}
where we have neglected an irrelevant phase.

We get seven contributions. 
The terms proportional to $\epsilon^2$ arise from the $a^3\epsilon^2 \zeta \dot \zeta^2$ term
\begin{equation}
2i\int_{-\infty}^{\tau_{end}} d\tau~ a^2 \epsilon^2
\left(\prod_i u_{i}(\tau_{end}) \right) 
\left( u^*_{1} \frac{du^*_{2}}{d\tau} \frac{du^*_{3}}{d\tau} +
{\rm two~perm} \right)
(2\pi)^3 \delta^3(\sum_i \bk_i) + {\rm c.c.} ~,
\label{eqn:term1}
\end{equation}
the $a \epsilon^2 \zeta (\partial \zeta)^2$ term
\begin{equation}
-2i \int_{-\infty}^{\tau_{end}} d\tau~ a^2 \epsilon^2
\left(\prod_i u_{i}(\tau_{end}) u^*_{i}(\tau) \right) 
\left(\bk_1\cdot \bk_2 + {\rm two~perm} \right)
(2\pi)^3 \delta^3(\sum_i \bk_i) + {\rm c.c.} ~,
\label{eqn:term2}
\end{equation}
and the $-2a\epsilon \dot\zeta (\partial\zeta)
(\partial\chi)$ term
\begin{equation}
-2i\int_{-\infty}^{\tau_{end}} d\tau~ a^2 \epsilon^2
\left(\prod_i u_{i}(\tau_{end}) \right) 
\left( u^*_{1} \frac{du^*_{2}}{d\tau} \frac{du^*_{3}}{d\tau}
\frac{\bk_1\cdot \bk_2}{k_2^2} + {\rm five~perm} \right)
(2\pi)^3 \delta^3(\sum_i \bk_i) + {\rm c.c.} ~.
\label{eqn:term3}
\end{equation}
The term proportional to $\epsilon \dot \eta$ is
\begin{equation}
i \left( \prod_i u_{i}(\tau_{end}) \right)  \int_{-\infty}^{\tau_{end}} d\tau a^2 \epsilon  \eta' 
\left( u_{1}^*(\tau) u_{2}^*(\tau) \frac{d}{d\tau} u_{3}^*(\tau)
+ {\rm two~perm} \right) (2\pi)^3 \delta^3(\sum_i \bk_i) + {\rm c.c.}
~.
\label{term4}
\end{equation}
The terms proportional to $\epsilon^3$ include
that from $\epsilon/2a \partial\zeta
\partial \chi \partial^2\chi$ term
\begin{equation}
\frac{i}{2} \int_{-\infty}^{\tau_{end}} d\tau~ a^2 \epsilon^3
\left(\prod_i u_{i}(\tau_{end}) \right) 
\left( u^*_{1} \frac{du^*_{2}}{d\tau} \frac{du^*_{3}}{d\tau}
\frac{\bk_1\cdot \bk_2}{k_2^2} + {\rm five~perm} \right)
(2\pi)^3 \delta^3(\sum_i \bk_i) + {\rm c.c.} ~,
\label{eqn:term5}
\end{equation}
and that from $\frac{\epsilon}{4a} (\partial^2\zeta)
(\partial\chi)^2$ term
\begin{equation}
\frac{i}{2} \int_{-\infty}^{\tau_{end}} d\tau~ a^2 \epsilon^3
\left(\prod_i u_{i}(\tau_{end}) \right) 
\left( u^*_{1} \frac{du^*_{2}}{d\tau} \frac{du^*_{3}}{d\tau}
k_1^2 \frac{\bk_2 \cdot \bk_3}{k_2^2 k_3^2} + {\rm two~perm} \right)
(2\pi)^3 \delta^3(\sum_i \bk_i) + {\rm c.c.} ~.
\label{eqn:term6}
\end{equation}
The field redefinition (\ref{FieldRedef}) contributes
\begin{equation}
\frac{\eta}{2} ~ |u_{2}|^2 |u_{3}|^2 \Bigg|_{\tau \to \tau_{end}}
(2\pi)^3 \delta^3(\sum_i \bk_i) + {\rm two~perm} ~.
\label{eqn:redeterm}
\end{equation}
In these equations, 
the ``two perm'' stands for two other terms that are symmetric under
permutations of the indices 1, 2 and 3.

In Ref.~\cite{Maldacena:2002vr,Seery:2005wm,Chen:2006nt}
where sharp features and non-attractor initial condition 
are absent, terms (\ref{eqn:term1}), (\ref{eqn:term2}), (\ref{eqn:term3}) and
(\ref{eqn:redeterm}) give the leading
contributions to non-Gaussianities. 
In Ref.~\cite{Chen:2006xj} 
where features are present on an otherwise flat
potential, term (\ref{term4}) is the leading term.  
For the non-attractor initial condition that we consider in
Sec.~\ref{Sec:IC}, the terms (\ref{eqn:term5}) and
(\ref{eqn:term6}) can also give important
contributions in addition to other terms, because
$\epsilon$ and $\eta$ can be large initially.

\section{Non-Gaussianities from initial conditions}
\label{App:IC} 

In this appendix, we give some details on the estimate of
non-Gaussianity from the 
fast-roll period discussed in Sec.~\ref{Sec:IC}.
In terms of the conformal time $\tau$, the scale
factor of the fast-roll period becomes
\bea
a = \frac{h_s}{H_s} \sqrt{1+2h_s \tau} ~, ~~~ \tau_i \le \tau \le 0
~.
\label{fastrolla}
\eea
Here the fast-roll starts at $\tau= \tau_i$ and end
at $\tau=0$.
$h_s$ is the conformal Hubble parameter $h\equiv a'/a$ evaluated
at the beginning the slow-roll inflation period (i.e.~the end of the
fast-roll period) $\tau =0$. $H_s$ is the corresponding Hubble parameter
in terms of $t$, $H =h/a$. This fast-roll period is immediately
connected to the inflation period 
in which the scale factor grows as
\bea
a = \frac{h_s/H_s}{1-h_s\tau} ~, ~~~ 0\le \tau < 1/h_s ~.
\eea

We first look at the power spectrum discussed in
\cite{Contaldi:2003zv}. 
Using (\ref{fastrolla}), one gets
\bea
2a^2 H^2 = \frac{2h_s^2}{(1+2h_s\tau)^2} ~,
\\
\epsilon \approx 3 ~, ~~~~~ \eta \approx 0 ~.
\eea
From the definition $z\equiv a \sqrt{2\epsilon}$, we get
\bea
\frac{z''}{z} \approx - \frac{h_s^2}{(1+2h_s\tau)^2} ~.
\eea
Picking a special initial condition for $v_k$, the equation of motion
$v_k'' +k^2 v_k - \frac{z''}{z} v_k =0$
can be solved as
\bea
v_k(\tau) = \sqrt{\frac{\pi}{8}} \frac{1}{\sqrt{h_s}}
(1+2h_s\tau)^{1/2} H^{(2)}_0 (k\tau + \frac{k}{2h_s}) ~,
\label{vfastroll}
\eea
where $H^{(2)}_0$ is the second Hankel function with index $0$. 
Requiring this solution and its first derivative to continuously match
to the slow-roll results
\bea
v_k(\tau) = C_1 e^{-i k\tau + i k/h_s} \left( 1+ \frac{i}{-k\tau +
k/h_s} \right) + C_2 e^{i k\tau - i k/h_s} \left( 1+ \frac{i}{k\tau -
k/h_s} \right)
\eea
at $\tau = 0$ determines the coefficients $C_1$ and $C_2$
\bea
C_1 = \sqrt{\frac{\pi}{32h_s}} e^{-i k/h_s} \left[ H^{(2)}_0 \left
( \frac{k}{2h_s} \right) - \left(\frac{h_s}{k} +i \right) H^{(1)}_1 \left
( \frac{k}{2h_s} \right) \right] ~,
\nonumber \\
C_2 = \sqrt{\frac{\pi}{32h_s}} e^{i k/h_s} \left[ H^{(2)}_0 \left
( \frac{k}{2h_s} \right) - \left(\frac{h_s}{k} -i \right) H^{(1)}_1 \left
( \frac{k}{2h_s} \right) \right] ~.
\eea
The power spectrum is 
\bea
P_k = \frac{H^2 {k}}{(2\pi)^2 \epsilon} |C_1-C_2|^2 ~.
\eea
At large scale for $k \ll 2h_s$,
\bea
P_k \to \frac{H^2 k^3}{(2\pi)^3 \epsilon h_s^3} \Bigg|\ln
\frac{k}{2h_s} \Bigg|^2 ~.
\eea
Hence we see that $P_k \to 0$ as $k\to 0$ resulting a suppression for
the large scales. For $k \gtrsim h_s$, it relaxes to the usual
attractor value $P_k = H^2/(8\pi^2\epsilon)$.

Now we estimate the order of magnitude of the contribution to 3-point
function by this fast-roll period.
As an example, we look at the $\epsilon^3$ terms (\ref{eqn:term5}) and
(\ref{eqn:term6}). The $\epsilon^2$ terms are expected to 
 be of a similar size.
We look at the modes $k\gtrsim h_s$ so the asymptotic value of $u_k$ at
the end of inflation $\tau_{end}$ is approximately that of the
slow-roll inflation, 
\bea
u_k(\tau_{end}) \approx \frac{i H_s}{\sqrt{4\epsilon_s k^3}} e^{-i k\tau}
~, 
\eea
as we see from the discussions in the last paragraph.

The net contribution of (\ref{eqn:term5}) and (\ref{eqn:term6}) to the 3pt
is
\bea
-\frac{7\pi^{9/2}}{64} \frac{H_s^4}{k^{9/2}h_s^{3/2}} \left
( \frac{\epsilon_f}{\epsilon_s} \right)^{3/2} \CR~
\delta^3 (\bk_1+\bk_2+\bk_3)  ~,
\eea
where
\bea
\CR = {\rm Re} \left[ \int_{\frac{k}{2h_s} \left(\frac{H_s}{H_f}
\right)^{2/3} } ^{\frac{k}{2h_s}} 
~dx ~x~ H_0^{(2)*} (x) H_1^{(2)*}(x) H_1^{(2)*}(x) \right ] ~.
\label{CRdef}
\eea
We have integrated from the start of the fast-roll period $\tau_i$ to
the beginning of the slow-roll period $\tau=0$. We have written 
the lower
limit of the integration in terms of the Hubble parameter at the
beginning of the fast-roll, $H_f$.

Comparing to the WMAP's ansatz in the limit $k_1=k_2=k_3$
\bea
(2\pi)^7 \frac{9}{10k^6} f_{fast} P_k^2 \delta^3 (\bk_1+\bk_2+\bk_3) ~,
\eea
we get a scale dependent amplitude for this fast-roll period
\bea
f_{fast}\approx -0.34 \left(\frac{k}{h_s}\right)^{3/2}
\epsilon_s^{1/2}
\epsilon_f^{3/2} \CR ~.
\label{fNLfastroll}
\eea
The integrand in (\ref{CRdef}) starts oscillating for large $x$. So
for very large $k$, $k \gg 2h_s \left({H_s}/{H_f}\right)^{2/3}$,
this integration approaches zero. Therefore, 
we look at the modes $k \lesssim 2h_s 
\left({H_s}/{H_f}\right)^{2/3}$. 
For example, for $k/h_s \approx 10$, $\epsilon_f\approx 3$,
$\epsilon_s \approx 0.01$, $H_s/H_f \approx 0.1$, we have
$\CR = \CO(1)$ and $f_{fast} \approx \CO(0.2)$. This is consistent
with the rough argument given in (\ref{fNLfast}).

\section{Slow-roll parameters in the resonance model}

\label{App:Res}

In this appendix, we work out some details on the behavior of the
slow-roll parameters in the resonant model. Using (\ref{Vexample}) as
the explicit example,  we decompose (and similarly for other variables)
\bea
\phi(t) = \phi_0(t) + \phi_{\rm osci} (t) ~,
\eea
where the $\phi_0$ is the unperturbed background evolution, and
$\phi_{\rm osci}$ is the oscillation caused by the small high
frequency ripples imposed on the potential.
Since the oscillation frequency $\omega \gg H$, the equation of
motion for the oscillatory part is
\bea
\ddot \phi_{\rm osci}  + (V')_{\rm osci} \approx 0 ~,
\label{EomOsci}
\eea
where the term $3H \dot \phi_{\rm osci}$ is much smaller than $\ddot
\phi_{\rm osci}$ and neglected.
The dominant contribution to $(V')_{\rm osci}$ comes from
\bea
V_{\rm osci} = \half m^2 \phi^2 c \sin \frac{\phi}{\Lambda} ~.
\label{Vosci}
\eea
Using the background evolution
\bea
\phi_0(t) = -\frac{\sqrt{6}}{3} m t + \phi_i ~,
\eea
we can solve (\ref{EomOsci}) and get
\bea
\phi_{\rm osci} \approx \frac{3}{4} c\phi_0^2 \Lambda \cos
(\frac{\phi_0(t)}{\Lambda}) ~.
\eea
With this result, one can check that (\ref{Vosci}) is indeed the
leading oscillatory term in (\ref{Vexample}).

We next calculate the oscillatory part of $H$ and $\dot H$.
Differentiating
\bea
3H^2 = V+\dot \phi^2/2 ~,
\label{Hsquare}
\eea
we get 
\bea
\dot H = -\frac{1}{2} \dot \phi^2 ~.
\eea
So, we have
\bea
\frac{\dot H_{\rm osci}}{\dot H_0} = -\frac{3c\phi_0^2}{2} \sin
(\frac{\phi_0(t)}{\Lambda}) ~.
\label{Hdotosci}
\eea
From (\ref{Hdotosci}), we get
\bea
\frac{H_{\rm osci}}{H_0} = \frac{3c \phi_0 \Lambda}{2\mpl^2} 
\cos (\frac{\phi_0(t)}{\Lambda}) ~.
\eea
Note that if we try to get $H_{\rm osci}$ 
from (\ref{Hsquare}), the leading
terms will cancel and the subleading terms are not reliable in our
approximation.

\begin{figure}[t]
\myfigure{5in}{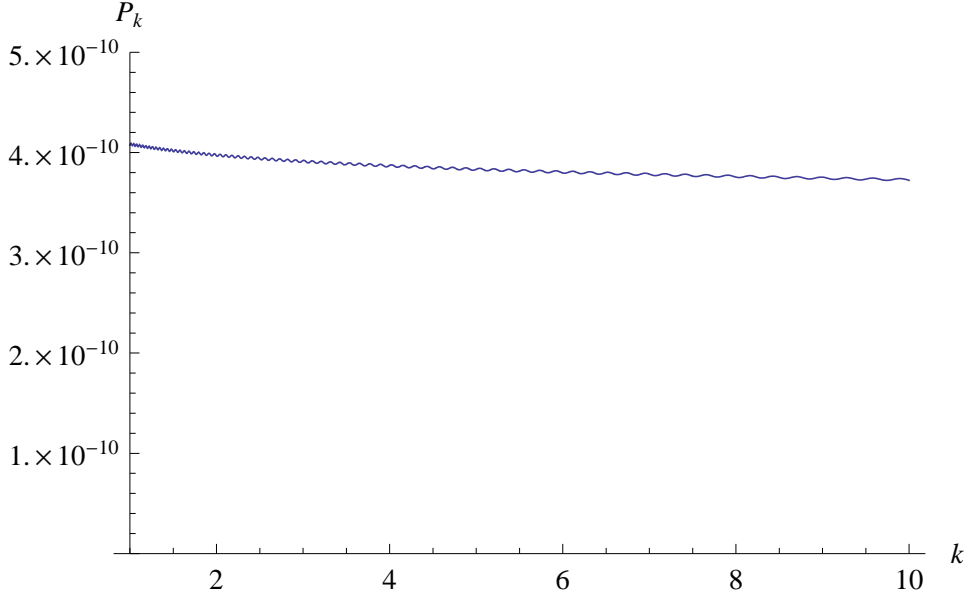}
\caption{The power spectrum $P_k$ for the resonance model. There is a
small $\CO(10^{-3})$ oscillation. 
The frequency (in $k$-space) is identical to that
of the 3-point correlation function (in $K$-space) in
Eq.~(\ref{eqn:resonanceansatz}). This is also different from
the sharp feature case where the frequency (in
$k$-space) of the power spectrum is twice of the frequency
(in $K$-space) of the 3-point correlation function \cite{Chen:2006xj}.}
\label{fig:resonancepower}
\end{figure}

Now we calculate the oscillatory part of the slow-roll parameters.
It is easy to see from the above results 
that the leading correction term for $\epsilon=-\dot H/H^2$
comes from that for $\dot H$, so
\bea
\epsilon_{\rm osci} &\approx& \epsilon_0 \frac{\dot H_{\rm osci}}{\dot
H_0}
\nonumber \\
&\approx& -3c \sin (\frac{\phi_0(t)}{\Lambda}) ~.
\eea
The leading correction term for $\eta=\dot \epsilon/(\epsilon H)$ 
comes from that for $\dot \epsilon$,
\bea
\eta_{\rm osci} &\approx& \eta_0 \frac{\dot \epsilon_{\rm
osci}}{\dot\epsilon_0} 
\nonumber \\
&\approx& \frac{3c\phi_0}{\Lambda} \cos (\frac{\phi_0(t)}{\Lambda}) \label{eqn:etaoscires}
~.
\eea
Taking the time derivative of the above equation (\ref{eqn:etaoscires}) we get
\bea
\dot \eta_{\rm osci} &\approx& \sqrt{6} \frac{cm\phi}{\Lambda^2} 
\sin (\frac{\phi_0(t)}{\Lambda}) ~,
\eea
which we can also derive directly by plugging in the potential Eq.~(\ref{Vexample}) into the definiton for $\eta$, Eq.~(\ref{eqn:eta}), and keeping the leading oscillatory term.

The most important term in the linear equation of motion
(\ref{quadeom}) is
\bea
\frac{z''}{z} = 2a^2 H^2 \left( 
1-\frac{\epsilon}{2} +\frac{3}{4}\eta
-\frac{1}{4}\epsilon \eta + \frac{1}{8} \eta^2 + \frac{1}{4}
\frac{\dot \eta}{H} \right)~.
\label{z''/z}
\eea
The most significant oscillating component comes from the last term
\bea
\frac{\dot \eta_{\rm osci}}{4H} = \frac{3c}{2\Lambda^2} 
\sin (\frac{\phi_0(t)}{\Lambda}) ~.
\label{doteta/H}
\eea
This term causes a dramatic oscillation of the freeze-out scale of the perturbation, which we
define to be $a\sqrt{z/z''}$. For example, in terms of the numerical
number (\ref{Nparameters}), the amplitude of (\ref{doteta/H}) is
$1.53$, of the same order of the non-oscillating component $1$.
The horizon size that we used in the main text refers to the averaged
non-oscillating component.
This oscillation affects the power spectrum, which we present
numerically in Fig.~\ref{fig:resonancepower}. The oscillation in the
power spectrum is
tiny, but is potentially observable if the fidelity of our CMB data is
very high. The structure of these oscillations is reminiscent of
those of the ``trans-Planckian'' models of inflation, but with the
distinctive scale-dependent oscillation frequency which has the same
properties as
we discussed at the end of Sec.~\ref{sect:subhorizon} for the
non-Gaussianity.
For relevant details on constraining such oscillatory power spectrum
on the CMB sky, see Ref.~\cite{Okamoto:2003wk,Easther:2005yr}.

\end{document}